\documentclass[sigconf]{acmart}
\AtBeginDocument{%
  }

\setcopyright{acmlicensed}
\copyrightyear{2025}
\acmYear{2025}
\acmDOI{XXXXXXX.XXXXXXX}
\acmConference[RecSys '25]{19th ACM Conference on Recommender Systems}{September 22-26,
  2025}{Prague, Czech Republic}

\usepackage{tcolorbox}
\usepackage{xcolor}
\usepackage{listings}
\usepackage{longtable}
\usepackage{booktabs}
\usepackage{colortbl} 
\usepackage{multirow}
\tcbset{
  promptbox/.style={
    colback=black!0,  
    coltext=black,    
    colframe=black,   
    arc=0mm,
    boxrule=1pt,      
    left=1mm,
    right=1mm,
    top=1mm,
    bottom=1mm,
  }
}

\begin{document}

\title{Grocery to General Merchandise: A Cross-Pollination Recommender using LLMs and Real-Time Cart Context}

\author{Akshay Kekuda\textsuperscript{*}, Murali Mohana Krishna Dandu\textsuperscript{*}, Rimita Lahiri\textsuperscript{*}, Shiqin Cai\textsuperscript{*} \\
Sinduja Subramaniam, Evren Korpeoglu, Kannan Achan \\
\textit{Walmart Global Tech, Sunnyvale, CA, USA}}

\renewcommand\footnotetextcopyrightpermission[1]{} 

\renewcommand{\shortauthors}{Akshay Kekuda, Murali Mohana Krishna Dandu, Rimita Lahiri, Shiqin Cai}

\begin{abstract}

Modern e-commerce platforms strive to enhance customer experience by providing timely and contextually relevant recommendations. However, recommending general merchandise to customers focused on grocery shopping – such as pairing milk with a milk frother – remains a critical yet under-explored challenge. This paper introduces a cross-pollination (XP) framework, a novel approach that bridges grocery and general merchandise cross-category recommendations by leveraging multi-source product associations and real-time cart context. Our solution employs a two-stage framework: (1) A candidate generation mechanism that uses co-purchase market basket analysis and LLM-based approach to identify novel item-item associations; and (2) a transformer-based ranker that leverages the real-time sequential cart context and optimizes for engagement signals such as add-to-carts. Offline analysis and online A/B tests show an increase of 36\% add-to-cart rate with LLM-based retrieval, and 27\% NDCG@4 lift using cart context-based ranker. Our work contributes practical techniques for cross-category recommendations and broader insights for e-commerce systems.

\end{abstract}

\maketitle

\begingroup
  \renewcommand{\thefootnote}{\fnsymbol{footnote}}
  \footnotetext[1]{These authors contributed equally to this work.}
\endgroup

\section{Introduction}

E-commerce platforms have revolutionized modern retail, offering customers a convenient way to shop for a wide range of products such as online groceries (eg. vegetables, meats) or general merchandise (eg. cookware, electronics). While customers frequently shop for Online Groceries (OG), they may also be interested in complementary General Merchandise (GM) that can facilitate and spark discovery shopping. For instance, a customer purchasing pet food might discover an interest in pet toys, or someone buying milk might appreciate a recommendation for a milk frother. Offline analysis revealed that users who shop across both OG and GM categories generate 2.5x more revenue than single-category shoppers. Inspired by cross-pollination in nature, where genetic diversity creates stronger hybrids, our approach bridges distinct shopping domains to create an enriched experience that makes cross-category discovery both seamless and engaging.

Traditional methods to cross-category recommendation, such as collaborative filtering and matrix factorization \cite{Zhou2019HelpfulnessAwareMF}, have been extended with deep learning techniques like Neural Collaborative Filtering (NCF) \cite{He2017NeuralCF} and knowledge graphs \cite{Wang2019KGATKG} to model complex relationships in cross categories. While these approaches have been effective in recommending similar or frequently co-purchased items, they often struggle with category bias, limiting recommendations within predefined domains. Cross-domain approaches often rely on deep learning \cite{Elkahky2015AMD} and transfer learning \cite{Pan2010TransferLI}\cite{Li2020ATLRecAA} to transfer knowledge across domains but typically rely on explicit user interactions between categories. Recent advances in Graph Neural Networks (GNN) \cite{Kanagawa2018CrossdomainRV} have improved cross-domain modeling, but they require extensive labeled data.  

The key contributions of this work include: 
1) Specific focus on Grocery to General Merchandise transition: We tackle the unique challenge of bridging fundamentally different shopping behaviors - from low-priced, routine grocery purchases to higher-priced, discretionary general merchandise items. 
2) LLM-Powered Candidate Generation and Evaluation: We leverage LLMs to generate product associations, a novel approach that can capture more nuanced, contextual, and potential multi-hop relationships between OG and GM items. We also propose a systematic way to evaluate the LLM-generated recommendations and the quality of retrieved items from these recommendations.
3) Real-Time Cart Context Utilization: Our system incorporates sequential cart context based ranker, allowing for highly dynamic and personalized recommendations.

\begin{figure*}[tp]
  \centering
  \includegraphics[width=\linewidth]{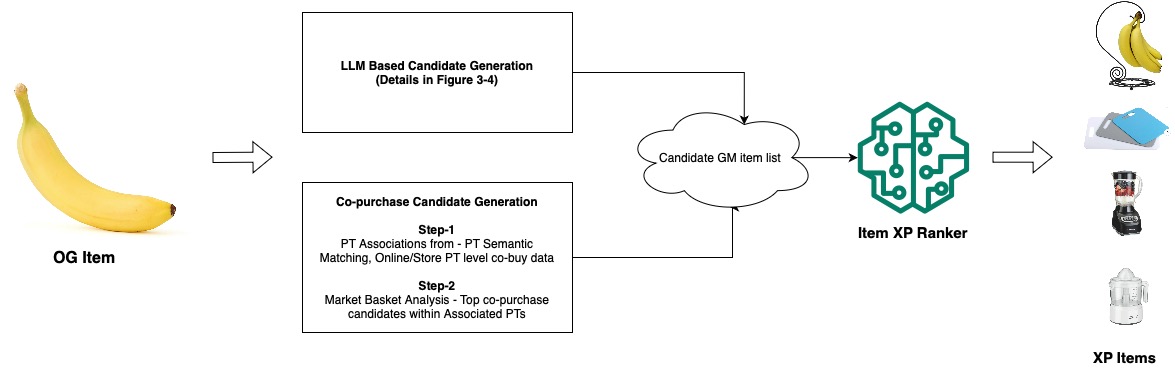}
  \caption{Item Cross-Pollination (XP) recommendation framework}
  \Description{Recommendation framework for item xp recommendations}
  \label{item_xp_fig}
\end{figure*}

\begin{figure*}[tp]
  \centering
  \includegraphics[width=\linewidth]{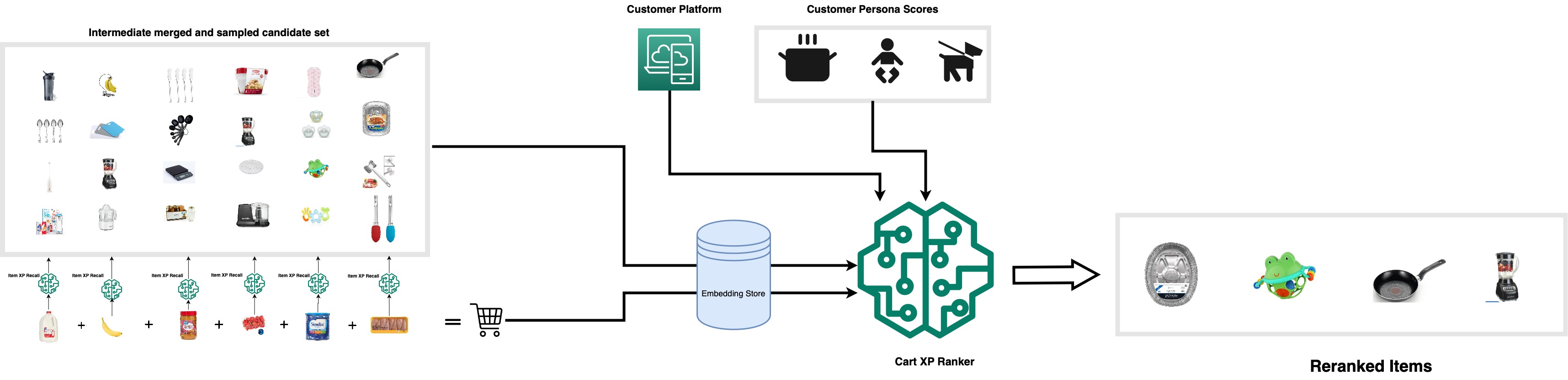}
  \caption{Cart Cross-Pollination (XP) recommendation framework}
  \Description{Recommendation framework for real time cart xp recommendations}
  \label{e2e_fig}
\end{figure*}

\section{Methodology}
In this section, we discuss the candidate generation process for grocery items (anchors), and the neural ranker that ranks recommendations based on real-time cart-context. Figures \ref{item_xp_fig}, \ref{e2e_fig} details our end to end recommendation framework.

\subsection{Item XP Recommendation Framework}

Our system employs two complementary approaches for candidate generation: a historical signals-based method that identifies and ranks cross-category relationships, and an LLM-based approach that generates contextually relevant recommendations beyond observed purchase patterns.

\subsubsection{Product-Type Associations \& Co-Purchase Based Candidates}
Our historical approach first establishes OG-to-GM product-type (PT) associations through multiple methods: PT name semantic matching and online/store PT-level co-purchase analysis. Once these PT-PT associations are identified, we select popular items within the associated GM product types as initial candidates. These candidates are then refined through market basket analysis (MBA) techniques \cite{agrawal1994discovery}.  

The lift metric, which quantifies the strength of association between items A and B, is computed as $\text{Lift}(A \rightarrow B) = \frac{P(A,B)}{P(A)P(B)}$, where $P(A,B)$ represents the joint probability of purchases within three-week time window to capture multi-session co-purchase shopping behavior. To ensure statistical significance, we apply minimum thresholds for support $(P(A,B))$ and confidence $(P(B|A))$ metrics.

\begin{figure*}[tp]
  \centering
  \includegraphics[width=\linewidth]{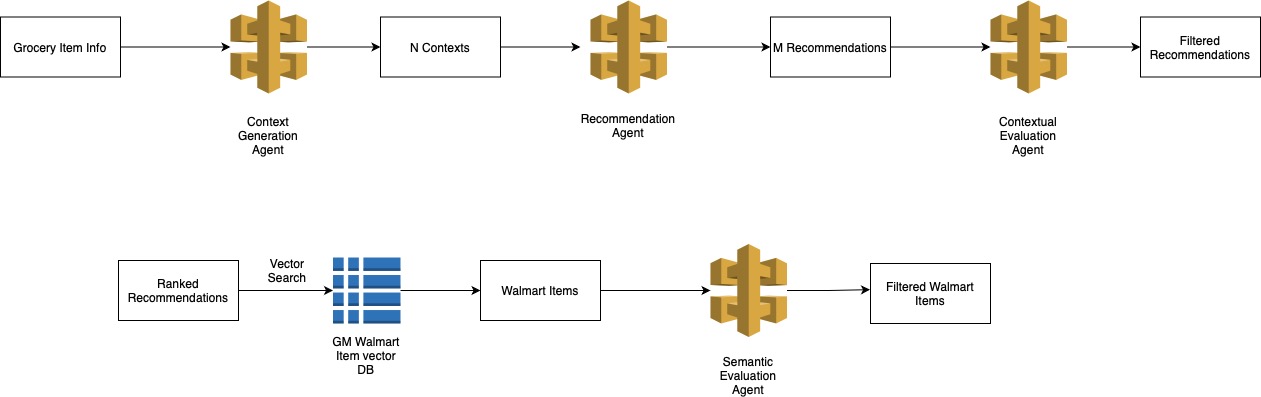}
  \caption{Illustration of Agentic LLM Recommendation and Scoring Framework}
  \Description{Illustration of Agentic LLM Recommendation and Scoring Framework}
  \label{xp_llm}
\end{figure*}

\begin{figure}[hbt!]
  \centering
  \includegraphics[width=\linewidth]{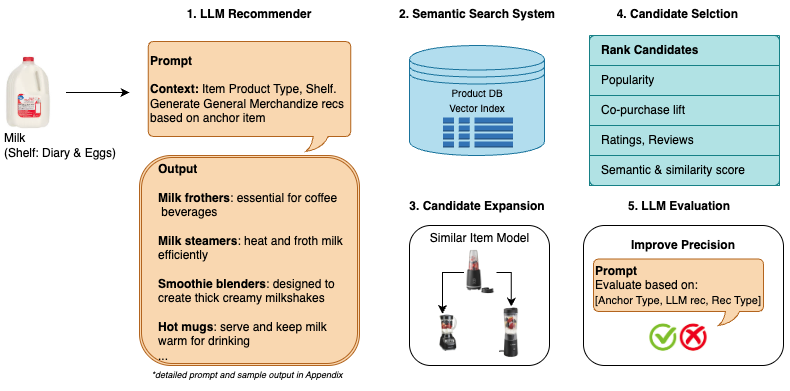}
  \caption{Illustration of LLM-Based Candidate Retriever and Evaluator}
  \Description{Illustration of LLM-Based Candidate Retriever and Evaluator}
  \label{llm_vis}
\end{figure}

\subsubsection{LLM-Based Candidate Generation and Evaluation}

While historical patterns provide valuable signals, they often reinforce existing behavior rather than encouraging category exploration. We leverage Large Language Models (LLMs) to identify novel, contextually relevant cross-category recommendations that might not emerge from historical data alone. 

We implement this through two complementary pipelines: an \textbf{agentic LLM recommendation pipeline} (Figure \ref{xp_llm}) for generating contextual recommendations, and a \textbf{retrieval-evaluation pipeline} (Figure \ref{llm_vis}) for matching and scoring these recommendations against our product catalog. We use gpt-4o \cite{openai2024gpt4o} as the LLM for all analyses, with detailed prompts provided in Appendix~\ref{appendix:llm-prompts}.

\paragraph{Agentic Recommendation Generation Pipeline}

Our recommendation generation follows a structured three-stage approach that transforms anchor grocery items into cross-category opportunities:

\textbf{Stage 1: Context Generation Agent.} Given an anchor grocery item, this agent generates thematic contexts that connect the anchor to potential cross-category opportunities. This stage focuses on identifying usage patterns, consumption contexts, and lifestyle scenarios associated with the anchor item (see Appendix~\ref{appendix:llm-prompts-theme}).

\textbf{Stage 2: Recommendation Agent.} For each generated theme, this agent produces specific product recommendations targeting non-grocery categories (cookware, home appliances, personal care, utilities). The LLM employs chain-of-thought reasoning to ensure recommendations are contextually relevant and cross-category focused (see Appendix~\ref{appendix:llm-prompts-theme-rec}).

\textbf{Stage 3: Contextual Evaluation Agent.} This agent evaluates each (anchor, recommendation, explanation) triplet using structured criteria including cross-category discovery potential, relevance to anchor item, explanation quality, and purchase likelihood. At this stage we filter out recommendations with a score $<$ 0.4 (see Appendix~\ref{appendix:llm-prompts-gen_evaluator}).

\paragraph{Semantic Search and Product Matching Pipeline}

The LLM recommendations serve as search queries mapped to our catalog via semantic search using the E5 embedding model \cite{e5model}. We generate catalog embeddings from item text and hierarchy metadata, bridging the semantic gap between LLM outputs and product representations. To enhance coverage and diversity, we expand candidate retrieval using an internal similar item model.

\paragraph{Dual Evaluation Framework}

The quality of retrieved items is evaluated using our \textbf{Semantic Evaluation Agent}, which employs two complementary scoring methods to ensure both semantic accuracy and contextual relevance:

\textbf{Method 1: LLM-As-Judge Evaluation.} This agent (see Appendix~\ref{appendix:llm-prompts-2}) scores each (anchor\_pt, llm\_rec, rec\_pt) triplet along four weighted dimensions, where:
\begin{itemize}
    \item \textbf{anchor\_pt:} The product type of the original anchor item
    \item \textbf{llm\_rec:} The LLM's textual recommendation description
    \item \textbf{rec\_pt:} The product type of the retrieved catalog item that matches the LLM recommendation
\end{itemize}

The evaluation dimensions below are chosen based on business alignment:
\begin{itemize}
    \item \textbf{Cross-Category Discovery (25\%):} The extent to which the recommendation comes from a meaningfully different, yet logically related category.
    \item \textbf{Relevance \& Coherence (35\%):} The logical, customer understandable connection between the anchor and recommendation.
    \item \textbf{Practical Utility (25\%):} The likelihood that the two items could be used or purchased together in realistic scenarios.
    \item \textbf{Matching Accuracy (15\%):} The degree to which the matched Walmart catalog item faithfully reflects the LLM's intended recommendation.
\end{itemize}

\textbf{Crucially, the evaluation is conducted at the product type (pt) level rather than the individual item level.} This design choice significantly reduces evaluation redundancy by avoiding repetitive assessments of items that share the same product type and exhibit similar recommendation patterns. By evaluating at the pt level, we capture the underlying semantic relationships and recommendation logic that generalize across item variants within the same product category, while dramatically improving computational efficiency and scalability. This agent produces a contextual \textbf{LLM Score} between 0 and 1 representing the contextual relevance of the retrieved item to the anchor item.

\textbf{Method 2: Cross-Encoder Evaluation.} In parallel, we measure semantic alignment using a Walmart-specific cross-encoder \cite{reimers2019sentence} model. This model takes as input the text of the LLM-generated recommendation and the candidate Walmart item, outputting a score between 0 and 1 representing recommendation relevance within the Walmart catalog. The cross-encoder accounts for catalog-specific semantics, including product type compatibility and functional relationships, ensuring high scores correspond to semantically coherent, commercially viable recommendations. This produces a \textbf{CE Score} between 0 and 1 representing semantic relevance.

\paragraph{Quality Assessment and Filtering}

For offline evaluation and filtering, we combine the contextual (LLM) and semantic (CE) scores into a single quality metric:

\begin{equation}
\text{Combined Score} = \text{CE Score} \times \text{LLM Score}
\end{equation}

This multiplicative approach ensures that low performance in either dimension heavily penalizes the final quality rating, allowing only recommendations that excel in both semantic alignment and contextual relevance to advance to the ranking stage.

Based on the combined score, recommendations are classified into five quality bands:
\begin{itemize}
    \item \textbf{Excellent (0.7–1.0):} High-quality recommendations ready for production use.
    \item \textbf{Very Good (0.6–0.7):} Strong recommendations with minor improvement opportunities.
    \item \textbf{Good (0.5–0.6):} Acceptable recommendations, requiring light manual review.
    \item \textbf{Fair (0.4–0.5):} Marginal recommendations, close to the filtering threshold.
    \item \textbf{Poor ($<$0.4):} Low-quality outputs removed before ranking.
\end{itemize}

\subsubsection{Human-in-the-Loop Validation}
We validated the LLM pipeline by manually reviewing recommendations for 200 diverse anchor items, achieving an overall \textbf{94\% relevancy rate}. The majority of observed issues stemmed from semantic matching challenges (70\%), with similar item expansion errors and LLM recommendation quality each contributing 15\%. 

Importantly, our human-in-the-loop evaluation demonstrated a \textbf{95\% alignment} between manual judgments and the combined scoring framework (Cross-Encoder $\times$ LLM-as-judge), validating the robustness of our automated evaluation methodology. The Combined Scoring system consistently filtered out irrelevant recommendations while preserving cross-category discovery potential, confirming its effectiveness for production deployment.

\subsection{Cart XP Recommendation Framework}
In Walmart’s e-commerce ecosystem, online grocery (OG) carts average $n \approx 20$ items and account for roughly $50\%$ of total order volume. Building effective cross-pollination (XP) recommendations at this stage is challenging because the cart evolves dynamically --- from $|C_t|$ to $|C_{t+1}|$ --- as shoppers add or remove products.

To address this, we extend the \textbf{Item XP} framework into the \textbf{Cart XP ranker}, a neural network--based system that re-scores GM candidates in real time based on the full, current cart context. The ranker is trained on six months of cart interaction logs, using only sessions where an OG item was in the cart and at least one GM item was viewed. Positive labels are defined as \emph{same-session clicks} that led to a GM add-to-cart within 7 days:
\[
P(\text{purchase} \mid \text{view}, \Delta t \leq 7 \text{ days})
\]
This label formulation rewards recommendations that drive not only immediate engagement but also short-term purchase intent.

The transition from Item XP to Cart XP enables a shift from \emph{single-item relevance} to \emph{whole-basket optimization}:
\begin{enumerate}
    \item \textbf{Item XP} provides the initial cross-category GM candidates for each OG product in the cart.
    \item \textbf{Cart XP} re-ranks these candidates by modeling the joint semantics and complementarity of all items currently in the cart, ensuring recommendations remain relevant as the basket evolves in real time.
\end{enumerate}

The model architecture implements a scoring function:
\begin{equation}
f(c, i) = \text{MLP}([\text{avg}(T(c)), A(i, c), e_p, e_u])
\end{equation}
where $T(c)$ denotes the transformer\cite{vaswani2017attention} encoding of cart items $c$, $A(i, c)$ represents the cross-attention\cite{vaswani2017attention} between GM item $i$ and cart items, $e_p$ denotes the platform embedding, and $e_u$ represents the customer feature vector. 

\subsubsection{Item Representation}
Each item's representation is computed as:
\begin{equation}
r_i = g([t_i, v_i, p_i, \tau_i])
\end{equation}
where $t_i$, $v_i$, $p_i$ and  $\tau_i$ represents the product title, product type, price and binary OG/GM item type embeddings with embedding sizes [768, 768, 16, 8] respectively. The title and product type embeddings are derived using the MPNet sentence transformer\cite{song2020mpnet}. To optimize model latency, item embeddings undergo transformation to a 128-dimensional space:
\begin{equation}
r_i' = W_r r_i, \quad W_r \in \mathbb{R}^{128 \times d}
\end{equation}

\subsubsection{Cart Transformer Encoder}
Cart representations are computed utilizing a transformer encoder. Given the cart item embeddings $X = [r_{c_1}, r_{c_2}, ..., r_{c_n}]$, the transformer encoder computes:
\begin{equation}
T(c) = \text{TransformerEncoder}(X)
\end{equation}
where the transformer layer comprises self-attention and feed-forward networks:

\begin{equation}
    \text{Self-Attention}(X) = \text{softmax} \left( \frac{(X W_Q)(X W_K)^T}{\sqrt{d_k}} \right) X W_V
\end{equation}

where $W_Q, W_K, W_V \in \mathbb{R}^{d \times d_k}$ represent learnable parameter matrices.
\subsubsection{Cross-Attention for Cart-Item Interaction}
To model interactions between the cart and candidate GM items, we employ a cross-attention mechanism. Given a candidate item embedding $r_i$ and cart embedding $T(c)$,

\begin{equation}
    A(i, c) = \text{softmax} \left( \frac{(r_i W_Q)(T(c) W_K)^T}{\sqrt{d_k}} \right) (T(c) W_V)
\end{equation}
where $W_Q, W_K, W_V \in \mathbb{R}^{d \times d_k}$ represent learnable parameter matrices.
This cross-attention module enables context-aware refinement of item relevance scores.

\subsubsection{Training Objective}

In this work, we optimize ranking performance using a \textit{list-wise softmax loss}\cite{cao2007learning}. This loss encourages the model to rank positively engaged items higher than negatively engaged ones in a list. Given the sets $\mathcal{P}$ and $\mathcal{N}$, the predicted scores are aggregated and transformed via a softmax function:
\begin{equation}
\mathcal{L}_{\text{softmax}} = - \frac{1}{|\mathcal{P}|} \sum_{i \in \mathcal{P}} \log \left( \frac{\exp(s_i / \tau)}{\sum_{j \in \mathcal{P} \cup \mathcal{N}} \exp(s_j / \tau)} \right)
\end{equation}
where $\tau$ is a temperature scaling factor that controls the sharpness of the probability distribution. The loss is computed as the negative log-likelihood of the positive item rankings. The loss operates within a mini-batch training framework, where it is averaged across all instances to compute the final optimization objective.

Appendix \ref{appendix:cart-transformer-complexity} details the complexity of the cart neural ranker and dimensions of the associated embeddings and features.

\section{Experiments}
\subsection{LLM Pipeline Quality Analysis}

This section presents a comprehensive evaluation of the proposed \textit{Theme-Based Recommendation} (\textbf{Theme+Rec}) pipeline against a baseline \textit{Naive Generation} approach. The Naive Generation approach entails prompting the LLM to directly provide cross category recommendations as provided in Appendix~\ref{appendix:prompt-naive-candidate} Our goal is to quantify not only the average performance improvements but also to explore the quality distribution, identify category-specific winners and laggards, uncover persistent failure modes, and examine how semantic and contextual judgments differ. The analyses are conducted over 4 most transacted grocery categories at Walmart - Food \& Beverages, Baby, Pet Supplies and Home \& Garden.

\subsubsection{Overall Quality Improvement}
Table~\ref{tab:llm_quality_comparison} compares average category level scores for both pipelines. Across all four evaluated product categories, Theme+Rec substantially outperforms Naive Generation, achieving an overall average score of \textbf{0.83} versus \textbf{0.69} for the baseline. The performance lift ranges from \textbf{+0.12} in \textit{Baby} to \textbf{+0.15} in \textit{Home \& Garden}, demonstrating that injecting contextual themes into the generation process yields consistent benefits.  

The largest gain in \textit{Pet Supplies} (+0.14) is particularly noteworthy given that this category contains many niche, functional products where generic recommendations often fail. The results suggest that theme-conditioning is especially valuable when item usage scenarios are diverse and require situational understanding.

\subsubsection{Score Distribution Analysis}
Table~\ref{tab:combined_distribution} reveals the distribution of combined scores across quality bands. More than half of the recommendations in every category are rated \textit{Excellent} ($>0.7$), with \textit{Food \& Beverages} leading at 61.60\%. This category also exhibits the lowest \textit{Poor} rate (9.57\%), indicating a particularly robust alignment between semantic content and contextual appropriateness.

The \textit{Pet Supplies} category, while showing significant improvement over baseline, still lags behind others in top-band performance (53.96\% Excellent) and has the highest proportion of \textit{Poor} scores (14.44\%). This points to persistent challenges in generating relevant, context-sensitive recommendations for highly specialized pet-related products.

\subsubsection{Category and Item-Level High Performers}
Table~\ref{tab:top_performers_combined} identifies high-performing item types across three metrics: CE score (semantic similarity), LLM score (contextual relevance), and their combined mean.  
\begin{itemize}
    \item \textbf{Semantic strength (CE):} Items such as \textit{Worcestershire Sauce}, \textit{Honey}, and \textit{Ground Coffee} exhibit near-perfect semantic alignment with their respective anchor contexts, with CE scores above 0.99.
    \item \textbf{Contextual strength (LLM):} Items like \textit{Olives}, \textit{Flours}, and \textit{Barbecue Sauces} achieve high contextual ratings ($>0.71$), showing that the model not only generated semantically related items but also matched them well to the intended usage scenarios.
    \item \textbf{Balanced excellence:} Items such as \textit{Fabric Starch} and \textit{Pet Grooming Wipes} rank among the top in combined scores, indicating both semantic and contextual strengths.
\end{itemize}

\subsubsection{Challenging Cases and Failure Modes}
While averages are high, certain items remain systematically difficult, as shown in Table~\ref{tab:poor_performers_combined}.  
\begin{itemize}
    \item \textbf{High CE failure rate:} \textit{Bird Food} and \textit{Fish Food} show the lowest semantic alignment scores (0.819 and 0.833 respectively), indicating fundamental semantic mismatches between pet food categories and general merchandise recommendations.
    \item \textbf{High LLM failure rate:} \textit{Pet Relaxants} demonstrates the poorest contextual evaluation (0.592), suggesting that while pet-related items may be semantically related, they fail to align with practical cross-category usage scenarios.
    \item \textbf{Dual-method struggles:} Products like \textit{Bird Food}, \textit{Pet Relaxants}, and \textit{Fish Food} consistently underperform across all metrics, achieving the lowest combined scores (0.520, 0.541, and 0.563 respectively).
\end{itemize}
These problematic items predominantly cluster in the \textit{Pet Supplies} category, suggesting that highly specialized pet products present unique challenges for cross-category discovery. The narrow applicability and limited thematic diversity of these products may require more targeted prompt engineering, specialized domain knowledge, or enhanced post-filtering mechanisms to improve recommendation quality.

\subsubsection{Semantic vs Contextual Disagreement}
Table~\ref{tab:score_differences} uncovers cases where semantic and contextual evaluations diverge sharply. Items such as \textit{Pretzels} and \textit{Pet Stain Remover} receive high CE scores (0.985 and 0.981 respectively) but comparatively lower LLM scores (0.645 and 0.644), resulting in the largest disagreement gaps (0.340 and 0.337). This indicates that while these items are lexically and conceptually aligned with their anchor categories, their practical fit to given usage scenarios is less convincing.

Conversely, items in the \textbf{lowest disagreement} category (e.g., \textit{Fish Food}, \textit{Bird Food}) show smaller gaps between evaluation methods (0.151 and 0.166 respectively), though both methods consistently rate these items poorly. This suggests fundamental challenges in cross-category discovery for highly specialized products where both semantic similarity and contextual relevance remain limited.

The \textbf{moderate disagreement} items (e.g., \textit{Baby Wipes}, \textit{Bananas}) with consistent ~0.275 score differences represent a middle ground where semantic associations provide reasonable foundation but contextual applicability requires further refinement. Understanding such disagreements can help diagnose where the model's semantic associations fail to translate into real-world relevance and guide targeted improvements in recommendation quality.

\subsubsection{Worst-Performing Recommendations}
Table~\ref{tab:worst_rec_all_scores} lists the five lowest-performing recommendations per category across all three scoring methods. Common failure patterns include:
\begin{enumerate}
    \item \textbf{Overly generic items} — e.g., \textit{Formula tracking app} for \textit{Baby}, which adds little practical value in context.
    \item \textbf{Misaligned purpose} — e.g., \textit{Airtight soup containers} in certain \textit{Food \& Beverages} scenarios where portability or serving utility was the priority.
    \item \textbf{Excessive niche specialization} — e.g., \textit{Automatic litter box cleaner}, which may be relevant but fails due to context specificity and low broad utility.
\end{enumerate}

\subsubsection{Metric-Specific Leaderboards}
To capture a more granular view of performance, Tables~\ref{tab:ce_mean_comprehensive}–\ref{tab:combined_mean_comprehensive} present top and bottom performers for each primary metric:
\begin{itemize}
    \item \textbf{CE Mean:} Semantic precision peaks with \textit{Worcestershire Sauce}, \textit{Disposable Diaper Bags}, and \textit{Straws}, while \textit{Bird Food}, \textit{Furnace Filters}, and \textit{Clams} occupy the bottom ranks.
    \item \textbf{LLM Mean:} Contextual alignment leaders include \textit{Shrimp \& Prawns}, \textit{Fabric Starch}, and \textit{Olives}. The weakest contextual fits occur in \textit{Pet Relaxants}, \textit{Applesauce \& Fruit Sauces}, and \textit{Matches}.
    \item \textbf{Combined Mean:} The highest-scoring items are \textit{Shrimp \& Prawns}, \textit{Fabric Starch}, and \textit{Mushrooms}, while \textit{Bird Food}, \textit{Pet Relaxants}, and \textit{Clams} trail behind.
\end{itemize}
This breakdown confirms that while many items are strong in both dimensions, some excel in one metric but falter in the other—insights that can guide model tuning priorities.

\subsubsection{Error Concentration by Poor-Score Percentages}
The \textbf{Pct Poor} metrics (Tables~\ref{tab:pct_poor_ce_comprehensive}–\ref{tab:pct_poor_combined_comprehensive}) highlight where the majority of low-quality outputs originate:
\begin{itemize}
    \item \textbf{CE Poor Concentration:} \textit{Diaper Pail Liners} (24.10\%), \textit{Bird Food} (23.65\%), and \textit{Furnace Filters} (16.08\%) are the most error-prone in semantic matching.
    \item \textbf{LLM Poor Concentration:} \textit{Pet Relaxants} (34.27\%), \textit{Applesauce \& Fruit Sauces} (33.06\%), and \textit{Matches} (31.05\%) exhibit the highest contextual rejection rates.
    \item \textbf{Combined Poor Concentration:} \textit{Pet Relaxants} (33.06\%), \textit{Bird Food} (31.95\%), and \textit{Diaper Pail Liners} (24.92\%) dominate the failure distribution, suggesting these are systemic problem cases.
\end{itemize}

\subsubsection{Business Impact Validation}

We validated the practical effectiveness of our LLM-based approach through coverage and engagement metrics during A/B testing:

\textbf{Coverage Impact:} LLM-based candidate generation achieved a 4.7x increase in unique recommendations (78.2k vs. 16.5k) compared to traditional association-based methods alone (Table \ref{tab:llm_recall_results}).

\textbf{Engagement Impact:} During dedicated A/B testing on item pages, we observed a 36\% lift in same-session GM add-to-cart rates (1.5\% vs. 1.1\%), demonstrating the practical effectiveness of LLM-generated cross-category recommendations in driving customer engagement beyond historically observed patterns.

The experimental evidence demonstrates that theme-based generation not only improves average quality scores but also reduces the proportion of low-quality outputs in most categories. The improvements are most pronounced for categories with high product diversity, where contextual grounding significantly boosts recommendation relevance.

\begin{table}[h]
\centering
\caption{LLM Pipeline Quality Comparison: Theme-Based vs Naive Generation}
\label{tab:llm_quality_comparison}
\small
\begin{tabular}{@{}lcc@{}}
\toprule
\textbf{Category} & \textbf{Naive} & \textbf{Theme+Rec} \\ 
\midrule
Food \& Beverages & 0.72 & 0.84 \\
Baby & 0.68 & 0.81 \\
Pet Supplies & 0.65 & 0.79 \\
Home \& Garden & 0.71 & 0.86 \\ 
\midrule
\textbf{Average} & \textbf{0.69} & \textbf{0.83} \\ 
\bottomrule
\end{tabular}
\end{table}

\begin{table}[h]
\centering
\caption{Combined Score Distribution by Category}
\label{tab:combined_distribution}
\resizebox{\columnwidth}{!}{%
\begin{tabular}{@{}lccccc@{}}
\toprule
\multirow{2}{*}{\textbf{Category}} & \textbf{Poor} & \textbf{Fair} & \textbf{Good} & \textbf{Very Good} & \textbf{Excellent} \\
& \textbf{(<0.4)} & \textbf{(0.4--0.5)} & \textbf{(0.5--0.6)} & \textbf{(0.6--0.7)} & \textbf{(0.7+)} \\
\midrule
Baby              & 11.87\% & 6.44\% & \textbf{6.76}\% & \textbf{18.63}\% & 56.30\% \\
Food \& Beverages & 9.57\%  & 5.11\% & 5.95\% & 17.77\% & \textbf{61.60}\% \\
Home \& Garden    & 10.35\% & 6.14\% & 6.63\% & 18.04\% & 58.84\% \\
Pet Supplies      & \textbf{14.44}\% & \textbf{7.15}\% & 6.50\% & 17.94\% & 53.96\% \\
\midrule
\textbf{Overall}  & \textbf{11.56\%} & \textbf{6.21\%} & \textbf{6.46\%} & \textbf{18.10\%} & \textbf{57.68\%} \\
\bottomrule
\end{tabular}%
}
\end{table}

\begin{table}[h]
\centering
\caption{Top Quality Anchors by Evaluation Method}
\label{tab:top_performers_combined}
\resizebox{\columnwidth}{!}{%
\begin{tabular}{c l c c c}
\toprule
\textbf{Category} & \textbf{Item Type} & \textbf{CE Score} & \textbf{LLM Score} & \textbf{Combined Score} \\
\midrule
\multicolumn{5}{l}{\textit{Highest Cross-Encoder Scores}} \\
Food \& Bev & Worcestershire Sauce & \textbf{0.999} & 0.655 & 0.654 \\
Food \& Bev & Honey & 0.995 & 0.684 & 0.681 \\
Food \& Bev & Ground Coffee & 0.992 & 0.681 & 0.676 \\
Home \& Garden & Straws & 0.989 & 0.671 & 0.664 \\
Pet Supplies & Pet Grooming Wipes & 0.967 & 0.694 & 0.671 \\
\midrule
\multicolumn{5}{l}{\textit{Highest LLM Evaluation Scores}} \\
Food \& Bev & Olives & 0.974 & \textbf{0.723} & 0.704 \\
Food \& Bev & Flours & 0.981 & 0.720 & 0.706 \\
Food \& Bev & Barbecue Sauces & 0.983 & 0.714 & 0.702 \\
Home \& Garden & Light Bulbs & 0.911 & 0.713 & 0.650 \\
Home \& Garden & Fabric Starch & 0.960 & 0.712 & 0.684 \\
\midrule
\multicolumn{5}{l}{\textit{Highest Combined Scores}} \\
Food \& Bev & Flours & 0.981 & 0.720 & \textbf{0.706} \\
Food \& Bev & Olives & 0.974 & 0.723 & 0.704 \\
Food \& Bev & Barbecue Sauces & 0.983 & 0.714 & 0.702 \\
Home \& Garden & Fabric Starch & 0.960 & 0.712 & 0.684 \\
Pet Supplies & Pet Grooming Wipes & 0.967 & 0.694 & 0.671 \\
\bottomrule
\end{tabular}%
}
\end{table}

\begin{table}[h]
\centering
\caption{Poor Quality Anchors by Evaluation Method}
\label{tab:poor_performers_combined}
\resizebox{\columnwidth}{!}{%
\begin{tabular}{c l c c c}
\toprule
\textbf{Category} & \textbf{Item Type} & \textbf{CE Score} & \textbf{LLM Score} & \textbf{Combined Score} \\
\midrule
\multicolumn{5}{l}{\textit{Lowest Cross-Encoder Scores}} \\
Pet Supplies & Bird Food & \textbf{0.819} & 0.653 & 0.520 \\
Pet Supplies & Fish Food & 0.833 & 0.682 & 0.563 \\
Pet Supplies & Small Animal Food & 0.852 & 0.669 & 0.579 \\
Baby & Diaper Pail Liners & 0.869 & 0.681 & 0.592 \\
Home \& Garden & Furnace Filters & 0.874 & 0.669 & 0.587 \\
\midrule
\multicolumn{5}{l}{\textit{Lowest LLM Evaluation Scores}} \\
Pet Supplies & Pet Relaxants & 0.915 & \textbf{0.592} & 0.541 \\
Food \& Beverages & Applesauce \& Fruit Sauces & 0.952 & 0.623 & 0.596 \\
Home \& Garden & Matches & 0.937 & 0.632 & 0.605 \\
Pet Supplies & Pet Diapers & 0.937 & 0.641 & 0.596 \\
Home \& Garden & Pet Stain Remover & 0.981 & 0.644 & 0.633 \\
\midrule
\multicolumn{5}{l}{\textit{Lowest Combined Scores}} \\
Pet Supplies & Bird Food & 0.819 & 0.653 & \textbf{0.520} \\
Pet Supplies & Pet Relaxants & 0.915 & 0.592 & 0.541 \\
Pet Supplies & Fish Food & 0.833 & 0.682 & 0.563 \\
Pet Supplies & Small Animal Food & 0.852 & 0.669 & 0.579 \\
Home \& Garden & Furnace Filters & 0.874 & 0.669 & 0.587 \\
\bottomrule
\end{tabular}%
}
\end{table}

\begin{table}[h]
\centering
\caption{Score Differences Between Cross-Encoder and LLM Evaluation Methods}
\label{tab:score_differences}
\resizebox{\columnwidth}{!}{%
\begin{tabular}{c l c c c}
\toprule
\textbf{Category} & \textbf{Item Type} & \textbf{CE Score} & \textbf{LLM Score} & \textbf{Score Difference} \\
\midrule
\multicolumn{5}{l}{\textit{Highest CE vs LLM Differences}} \\
Food \& Beverages & Pretzels & 0.985 & 0.645 & \textbf{0.340} \\
Home \& Garden & Pet Stain Remover & 0.981 & 0.644 & 0.337 \\
Food \& Beverages & Worcestershire Sauce & 0.999 & 0.669 & 0.330 \\
Food \& Beverages & Applesauce \& Fruit Sauces & 0.952 & 0.623 & 0.329 \\
Food \& Beverages & Vinegars & 0.982 & 0.654 & 0.328 \\
\midrule
\multicolumn{5}{l}{\textit{Moderate CE vs LLM Differences}} \\
Baby & Baby Wipes & 0.968 & 0.692 & 0.276 \\
Food \& Beverages & Bananas & 0.971 & 0.695 & 0.276 \\
Food \& Beverages & Apples & 0.977 & 0.702 & 0.275 \\
Home \& Garden & Paper Towels & 0.982 & 0.707 & 0.275 \\
Home \& Garden & Mop Heads & 0.964 & 0.689 & 0.275 \\
\midrule
\multicolumn{5}{l}{\textit{Lowest CE vs LLM Differences}} \\
Food \& Beverages & English Muffins & 0.908 & 0.704 & 0.204 \\
Baby & Diaper Pail Liners & 0.869 & 0.681 & 0.188 \\
Pet Supplies & Small Animal Food & 0.852 & 0.669 & 0.183 \\
Pet Supplies & Bird Food & 0.819 & 0.653 & 0.166 \\
Pet Supplies & Fish Food & 0.833 & 0.682 & \textbf{0.151} \\
\bottomrule
\end{tabular}%
}
\end{table}

\begin{table}[h]
    \centering
    \caption{Top 5 Worst Performing Recommendations per Category by CE, LLM, and Combined Scores (Min 20 instances)}
    \label{tab:worst_rec_all_scores}
    \resizebox{\columnwidth}{!}{%
    \begin{tabular}{c c c c}
    \toprule
    \textbf{Category} & \textbf{CE Score} & \textbf{LLM Score} & \textbf{Combined Score} \\
    \midrule
    \multirow{5}{*}{Baby} 
        & Baby food storage racks (0.000) & Silicone bib (0.100) & Baby food storage racks (0.000) \\
        & Formula scoop leveler (0.000) & Bedtime alarm clock (0.200) & Formula scoop leveler (0.000) \\
        & Formula tracking app (0.000) & Collapsible picnic mat (0.200) & Formula tracking app (0.000) \\
        & Gas relief drops dispenser (0.000) & Insulated food pouch carrier (0.217) & Gas relief drops dispenser (0.000) \\
        & Nipple flow testers (0.075) & Small microwave steamer (0.220) & Nutritional tracking app (0.027) \\
    \midrule
    \multirow{5}{*}{Food \& Beverages} 
        & Patty storage containers (0.100) & Airtight soup containers (0.125) & Patty storage containers (0.075) \\
        & Recliner cup holders (0.141) & Seatback organizers (0.225) & Recliner cup holders (0.101) \\
        & Burger seasoning shaker (0.300) & Tabletop organizers (0.225) & Airtight soup containers (0.125) \\
        & Calorie tracking app subscription (0.350) & Wireless meat thermometer (0.250) & Burger seasoning shaker (0.215) \\
        & Garlic butter spreaders (0.367) & Mini trash cans (0.275) & Breakfast station organizer (0.225) \\
    \midrule
    \multirow{5}{*}{Home \& Garden} 
        & Laundry chute systems (0.033) & Solar pool cover (0.100) & Laundry chute systems (0.022) \\
        & Absorbent floor pads (0.100) & Fabric wrinkle sprays (0.150) & Absorbent floor pads (0.072) \\
        & Anti-freezer burn wraps (0.100) & Dough proofing baskets (0.200) & Anti-freezer burn wraps (0.075) \\
        & Bathroom cleaning checklist (0.150) & Furniture moving sliders (0.200) & Solar pool cover (0.100) \\
        & Descaling solution dispenser (0.200) & Allergen-proof pillow covers (0.250) & Descaling solution dispenser (0.120) \\
    \midrule
    \multirow{5}{*}{Pet Supplies} 
        & Automatic litter box cleaner (0.000) & Backup manual feeder (0.150) & Automatic litter box cleaner (0.000) \\
        & Backup manual feeder (0.000) & Color-coded pet collars (0.150) & Backup manual feeder (0.000) \\
        & Cat feeding journal (0.000) & Slow feeder mat (0.158) & Cat feeding journal (0.000) \\
        & Chew monitoring timer (0.000) & Collapsible water dispenser (0.233) & Chew monitoring timer (0.000) \\
        & Dog mouthwash dispenser (0.000) & Color-coded feeding mats (0.233) & Dog mouthwash dispenser (0.000) \\
    \bottomrule
    \end{tabular}
    }
\end{table}


\begin{table}[h]
\centering
\caption{Evaluation of LLM based retrieval}
\label{tab:llm_recall_results}
\begin{tabular}{lcc}
\hline
\textbf{Method}          & \textbf{Novelty} & \multicolumn{1}{l}{\textbf{ATCR}} \\ \hline
Without LLM based Recall & 16.5k            & 1.1\%                             \\
With LLM based Recall    & 78.2k              & 1.5\%                             \\ \hline
                         & \textbf{+4.7x}   & \textbf{+36\%}                    \\ \hline
\end{tabular}
\end{table}

\subsection{Cart Neural Ranker Evaluation}

The cart-based ranker is evaluated against the existing production model deployed on cart pages. On the cart page we use the item XP pipeline to generate candidates for items in the cart. The production baseline employs heuristic ranking to rank this aggregated recall set based on customer persona scores and assigns importance to cart product types based on their distinct item counts, regardless of sequence or context. 

\begin{table}[h]
  \caption{Comparison of NDCG Lift across different models}
  \label{tab: ndcg gain results}
  \centering
  \resizebox{\columnwidth}{!}{%
  \begin{tabular}{l c c c c}
    \toprule
    \textbf{Experiment} & \textbf{Loss} & \multicolumn{3}{c}{\textbf{NDCG Lift \%}} \\
    \cmidrule(l){3-5}
    & & \textbf{@2} & \textbf{@4} & \textbf{@6} \\
    \midrule
    Production System & N/A & N/A & N/A & N/A\\
    DeepFM & Pairwise & +22.43 & +22.80 & +12.39\\
    Identity Cart Encoder & Pairwise & +25.05 & +25.29 & +14.39\\
    Bidirectional LSTM Cart Encoder & Pairwise & +24.60 & +25.13 & +14.28 \\
    Transformer Cart Encoder & Pairwise & +25.58 & +25.55 & +14.64 \\
    Transformer Cart Encoder + Cross Attention & Pairwise & +26.24 & +26.06 & +14.97\\
    \textbf{Transformer Cart Encoder + Cross Attention} & Listwise & \textbf{+27.19} & \textbf{+26.50} & \textbf{+15.34}\\
    \bottomrule
  \end{tabular}%
  }
\end{table}

We report NDCG metrics \cite{jarvelin2002cumulated} in Table \ref{tab: ndcg gain results} with various model architectures and loss functions. We first start our experiments using a pair-wise hinge loss \cite{burges2005learning} as our training objective. We keep a margin of $\delta=5$ for hinge loss in all our experiments. With a simple identity cart encoder we see a 25.05\% lift in NDCG@2 and a 25.29\% lift in NDCG@4 metrics. When we used a Bidirectional LSTM \cite{hochreiter1997long} to model the cart, we only see a lift of 24.60\% and 25.55\% lift in NDCG@2 and NDCG@4. Upon using a transfomer to represent the cart, we see 25.58\% and 25.55\% lift in NDCG@2 and NDCG@4. Using the cross attention layer further provides a 26.24\% and 26.06\% lift in NDCG@2 and NDCG@4. Finally, using a list wise softmax loss provides the best lift in NDCG metrics compared to our production system. We use a temperature of $\tau=5$ for our experiments.

\subsection{Production Deployment}
\subsubsection{LLM-Based Candidate Generation Deployment Pipeline}

We deploy the LLM-based candidate generation system to cover the 100k most transacted grocery items in our catalog. The initial generation pipeline required 24 hours for complete recommendation generation across all anchor items, followed by 12 hours for comprehensive evaluation using our dual-scoring framework. This one-time batch process creates a comprehensive mapping of cross-category recommendations that significantly expands our candidate pool beyond historical co-purchase patterns.

To maintain currency with catalog changes and seasonal variations, we execute an incremental pipeline weekly to generate recommendations for newly introduced grocery items and refresh existing recommendations based on updated performance metrics. This weekly refresh cycle ensures that our cross-category recommendations remain relevant as customer preferences and product availability evolve.

The generated recommendations are stored in our distributed recommendation cache with sub-100ms retrieval latency, enabling seamless integration with our real-time cart-based ranking system. Quality thresholds established during offline evaluation automatically filter low-scoring recommendations before they reach the production ranking stage.

\subsubsection{Cart-Based Ranker Deployment}

This model is deployed on the cart page to show relevant XP recommendations given OG anchor items added to customer's cart. As shown in Figure \ref{e2e_fig}, when a customer adds an OG item to their cart, the system retrieves the top-30 recommended items from the item-level XP model, and gets added to current candidate pool. Once the candidate pool reaches a size of 300, stratified sampling is applied across different product types to ensure diversity in the final candidate selection.

We store all our item embeddings in an embedding store that is hosted alongside the ranker model enabling real-time embedding retrieval. During every add-to-cart event, the system retrieves the cart item embeddings, XP candidate embeddings, customer persona scores, and platform details in real time. These are then fed to ranker, and results are diversified before finally displaying on the carousel. Our architecture is optimized for low-latency inference for an average 600 cart requests per second, enabling seamless updates to recommendations as the cart evolves and \textbf{providing dynamic, personalized, and scalable grocery basket recommendations} 

\section{Conclusion}
In this paper, we introduce a novel cross-pollination recommendation framework designed to bridge the gap between grocery and general merchandise and significantly promoting cross-category discovery. Our approach innovatively leverages market basket analysis and LLMs in candidate generation and utilizes real-time cart context for personalized session-based recommendations. Persistent weaknesses with the Agentic LLM generation pipeline cluster around specialized, narrow-application products where both semantic and contextual alignment are inherently challenging. Addressing these gaps will likely require:
\begin{enumerate}
    \item More domain-specific fine-tuning on niche product contexts.
    \item Enhanced negative sampling to penalize generic or tangential outputs.
    \item Post-generation filtering that jointly considers semantic proximity and context-specific utility.
\end{enumerate}
By focusing refinement efforts on these problem areas, future iterations of the model can further close the performance gap, particularly in \textit{Pet Supplies} and specialized \textit{Baby} product recommendations.. Additionally, we plan to enhance personalization using contextual signals such as purchase frequency, seasonal trends and user intent. 

\newpage

\bibliographystyle{ACM-Reference-Format}
\bibliography{xp-ref}

\newpage
\onecolumn  

\appendix  

\section{LLM Prompts}
\label{appendix:llm-prompts}

\subsection{Naive Candidate Generator}
\label{appendix:prompt-naive-candidate}
\begin{tcolorbox}[promptbox]
\begin{lstlisting}[backgroundcolor=\color{black!0}, basicstyle=\small\ttfamily\color{black}]
You are an e-commerce recommendation agent.
Recommend product types that are very specifically used with the anchor_item quoted in triple brackets.
anchor_item: {{item_input}}.

Thought process:
First, think about anchor item/type's common uses, consumption methods, related recipes, storage and preparation requirements, nutritional aspects, user personas and life styles. Next, based on this context, suggest non-grocery items.

Your recommendations should be from Cookware, Home Appliances, Personal Care and general house utilities categories.
Your recommendations should not include grocery related categories.
Do not include recommendations that are not directly used with given anchor_item.
We are looking for item types that are very specific to the given anchor_item.

Examples:
- For Bananas, recommendations could be Fruit baskets (User who buys Bananas and may buy other fruits would need a fruit basket to keep them). Others could be Banana Hangers, Banana Smoothie Makers etc.
- For Milk, it could be Milk Frothers (User who buys Milk would use it to make Coffee etc. which require Milk Frothers), Others could be Smoothie Blenders, Hot Mugs etc.

Your output must be in json format with following keys:
- recs: A Python list of top {n} recommendations.
  Your recommendations should not be detailed items but a broader item type of around 2-5 words.
- explanation: A Python list of short explanations for why you recommended that item type.

Do not output anything else except for the above json.
\end{lstlisting}
\end{tcolorbox}

\subsection{Theme Generator}
\label{appendix:llm-prompts-theme}
\begin{tcolorbox}[promptbox]
\begin{lstlisting}[backgroundcolor=\color{black!0}, basicstyle=\small\ttfamily\color{black}]
    Analyze the following item: {anchor_item} and identify the top 5 most popular usage contexts where this item is commonly used or consumed. Focus on:

    1. Most frequent usage occasions, routines, or activities
    2. Popular application methods or scenarios that drive purchase decisions
    3. Common household, personal, or lifestyle situations where this item is essential
    4. Peak usage times or situations (daily routines, maintenance schedules, special occasions, seasonal needs)

    Prioritize contexts that represent the highest volume usage patterns and mainstream consumer behavior.

    Examples:

    **Eggs:**
    1. **Breakfast** - Eggs are a staple breakfast protein, commonly prepared as scrambled, fried, poached, or in omelets
    2. **Baking** - Essential ingredient in cakes, cookies, breads, and pastries for binding, leavening, and structure
    3. **Protein Source** - High-quality complete protein for fitness enthusiasts, meal prep, and nutritious cooking
    4. **Holiday Cooking** - Traditional ingredient for deviled eggs, egg salad, and festive dishes during gatherings and celebrations
    5. **Quick Dinners** - Versatile protein for fast weeknight meals like fried rice, pasta carbonara, and simple egg-based dishes

    **Milk:**
    1. **Cereal** - Primary liquid component for breakfast cereals, providing essential pairing for morning meals
    2. **Beverages** - Consumed as standalone drink or mixed into coffee, tea, and flavored drinks for creaminess
    3. **Cooking** - Essential ingredient in baking, sauces, soups, and creamy dishes for texture and richness
    4. **Children's Nutrition** - Daily calcium and protein source for growing children, often consumed at meals and bedtime
    5. **Post-Workout Recovery** - High-quality protein beverage for muscle recovery after exercise and fitness activities

    **Dog Food:**
    1. **Daily Feeding** - Primary nutrition source for dogs, typically served twice daily in measured portions
    2. **Training Rewards** - Used as high-value treats during obedience training and behavioral reinforcement sessions
    3. **Health Management** - Specialized nutrition for weight control, allergies, and age-specific dietary needs
    4. **Emergency Preparedness** - Essential stockpiling for natural disasters, power outages, and unexpected supply chain disruptions
    5. **Multi-Pet Households** - Bulk purchasing and feeding coordination for households with multiple dogs of different sizes and ages
\end{lstlisting}
\end{tcolorbox}

\label{appendix:llm-prompts-theme}
\begin{tcolorbox}[promptbox]
\begin{lstlisting}[backgroundcolor=\color{black!0}, basicstyle=\small\ttfamily\color{black}]
    **Cotton Swabs:**
    1. **Personal Hygiene** - Daily ear cleaning and personal grooming routines for individual cleanliness
    2. **Makeup Application** - Precision tool for cosmetic touch-ups, blending, and detailed beauty work
    3. **Household Cleaning** - Small area cleaning for electronics, jewelry, and hard-to-reach spaces
    4. **Arts and Crafts** - Precision tool for painting details, applying adhesives, and creating texture in artistic projects
    5. **First Aid Care** - Medical application for wound cleaning, antiseptic application, and precise healthcare tasks

    **Baby Wipes:**
    1. **Diaper Changes** - Essential cleaning during infant diaper changes for hygiene and comfort
    2. **Quick Cleanups** - Immediate cleaning of spills, sticky hands, and messes throughout the day
    3. **Travel Care** - Portable cleaning solution for on-the-go infant and toddler care needs
    4. **Adult Personal Care** - Convenient hygiene solution for elderly care, bedridden patients, and personal freshening
    5. **Surface Sanitizing** - Quick disinfection of high-touch surfaces, toys, and equipment during illness prevention

    **Paper Plates:**
    1. **Party Events** - Disposable serving solution for birthdays, gatherings, and large group entertaining
    2. **Outdoor Dining** - Convenient dishware for picnics, camping, and backyard barbecues
    3. **Quick Meals** - Time-saving option for busy households avoiding dishwashing during hectic periods
    4. **Office Events** - Workplace celebrations, meetings, and catered events requiring easy cleanup and disposal
    5. **Emergency Situations** - Essential backup dishware during power outages, water restrictions, and natural disasters

    Output format: Python list with exactly 5 items in format:
    ['Context - brief explanation (5-6 words)', 'Context - brief explanation (5-6 words)', 'Context - brief explanation (5-6 words)']
\end{lstlisting}
\end{tcolorbox}

\subsection{Theme Recommendation Generator}
\label{appendix:llm-prompts-theme-rec}

\begin{tcolorbox}[promptbox]
\begin{lstlisting}[
  backgroundcolor=\color{black!0},
  basicstyle=\small\ttfamily\color{black},
  breaklines=true,
  postbreak=\mbox{\textcolor{red}{$\hookrightarrow$}\space}
]
    Based on the following popular usage contexts for {anchor_item}:

    {anchor_contexts}

    For each context, generate 10 complementary non-grocery item recommendations that are:
    - Essential tools/equipment that enhance the usage experience in that specific context
    - Items that solve common problems when using {anchor_item} in that context
    - Products that make the context more convenient, efficient, or enjoyable
    - Supporting items that improve the effectiveness or results when using {anchor_item}

    Focus on items from these categories: kitchen tools, home appliances, storage solutions, serving equipment, preparation tools, household utilities, automotive accessories, personal care tools, baby care products, pet care accessories, office supplies, cleaning equipment, maintenance tools, and organization products.

    Examples:

    **Eggs - Breakfast Context:**
    1. **Non-stick egg pan** - Specialized pan with optimal size and coating for perfect egg cooking without sticking
    2. **Egg ring molds** - Creates perfectly round fried or poached eggs for restaurant-quality breakfast presentation
    3. **Silicone egg cookers** - Microwave-safe containers for quick, mess-free egg preparation during busy mornings
    4. **Breakfast serving plates** - Designed with compartments to separate eggs from other breakfast items like toast
    5. **Egg separator tools** - Efficiently separates yolks from whites for recipes requiring specific egg components
    6. **Toaster** - Quick and even toasting of bread, bagels, and English muffins to accompany eggs
    7. **Breakfast trays** - Convenient serving trays for carrying eggs and sides from kitchen to dining area
    8. **Egg slicers** - Creates uniform slices of hard-boiled eggs for salads, sandwiches, and garnishes
    9. **Whisk and spatula set** - Essential tools for scrambling eggs and flipping fried eggs with ease
    10. **Salt and pepper shakers** - Essential condiments for seasoning eggs to taste during breakfast

    **Milk - Cereal Context:**
    1. **Cereal bowls** - Deep, wide bowls specifically designed to hold cereal and milk without spillage
    2. **Milk pitchers** - Pour-controlled containers for transferring milk from gallon jug to cereal bowls
    3. **Cereal storage containers** - Airtight containers that keep cereal fresh while milk is consumed
    4. **Breakfast serving trays** - Organized trays for carrying cereal, milk, and utensils from kitchen
    5. **Cereal spoons** - Deep-bowled spoons designed for optimal cereal and milk consumption
    6. **Milk frothers** - Creates frothed milk for cereal toppings, enhancing breakfast experience
    7. **Cereal dispensers** - Bulk storage and dispensing systems that keep cereal fresh and easily accessible
    8. **Refrigerator organizers** - Storage solutions that keep milk easily accessible and organized in the fridge
\end{lstlisting}
\end{tcolorbox}

\label{appendix:llm-prompts-theme-rec}
\begin{tcolorbox}[promptbox]
\begin{lstlisting}[backgroundcolor=\color{black!0}, basicstyle=\small\ttfamily\color{black}]
    9. **Breakfast placemats** - Easy-to-clean mats that protect surfaces from spills during cereal consumption
    10. **Milk storage containers** - Reusable containers for storing leftover milk after pouring into cereal bowls

    **Dog Food - Daily Feeding Context:**
    1. **Elevated dog bowls** - Raised feeding stations that improve digestion and reduce neck strain during eating
    2. **Food storage containers** - Airtight containers that keep dry dog food fresh and pest-free
    3. **Measuring cups** - Precise portion control tools for consistent daily feeding amounts and weight management
    4. **Feeding mats** - Waterproof mats that protect floors from spills and make cleanup easier
    5. **Automatic feeders** - Timed dispensers for consistent feeding schedules when owners are away
    6. **Dog food scoops** - Specialized scoops for easy and mess-free portioning of dry dog food
    7. **Water fountains** - Continuous water supply systems that encourage hydration alongside daily feeding
    8. **Dog food toppers** - Flavor enhancers that make dry food more appealing and nutritious for daily meals
    9. **Pet food storage bins** - Large-capacity containers that keep bulk dog food fresh and organized
    10. **Dog feeding stations** - All-in-one setups that include bowls, storage, and mats for complete feeding solutions

    **Cotton Swabs - Personal Hygiene Context:**
    1. **Bathroom organizers** - Storage solutions that keep cotton swabs clean, dry, and easily accessible
    2. **Magnifying mirrors** - Enhanced visibility tools for precise personal grooming and hygiene tasks
    3. **Travel containers** - Portable cases for cotton swabs during travel and on-the-go hygiene needs
    4. **Vanity lighting** - Improved illumination for detailed personal care and grooming activities
    5. **Disposal containers** - Hygienic waste receptacles specifically for used personal care items
    6. **Cotton swab holders** - Decorative and functional containers that keep swabs organized and within reach
    7. **Ear wax removal kits** - Comprehensive kits that include cotton swabs and other ear hygiene tools
    8. **Ear cleaning kits** - Comprehensive kits that include cotton swabs and other ear hygiene tools
    9. **First aid kits** - Essential medical supplies that include cotton swabs for wound care and cleaning
    10. **Personal grooming kits** - Multi-purpose kits that combine cotton swabs with other grooming essentials

    **Baby Wipes - Diaper Changes Context:**
    1. **Changing pad covers** - Waterproof, easy-clean surfaces that work perfectly with baby wipes for hygiene
    2. **Diaper caddies** - Organized storage that keeps wipes accessible during diaper changing sessions
    3. **Wipe warmers** - Heating devices that make cold wipes more comfortable for sensitive baby skin
    4. **Diaper pails** - Odor-sealing waste containers that work with wipes for complete diaper disposal
    5. **Portable changing mats** - Travel-friendly surfaces that pair with wipes for on-the-go diaper changes
\end{lstlisting}
\end{tcolorbox}

\label{appendix:llm-prompts-theme-rec}
\begin{tcolorbox}[promptbox]
\begin{lstlisting}[backgroundcolor=\color{black!0}, basicstyle=\small\ttfamily\color{black}]
    6. **Baby lotion dispensers** - Convenient bottles for applying lotion after using wipes during diaper changes
    7. **Wipe dispensers** - Easy-to-use containers that keep wipes moist and accessible during diaper changes
    8. **Baby care kits** - Comprehensive sets that include wipes, creams, and other essentials for diapering
    9. **Changing table organizers** - Storage solutions that keep wipes and other changing supplies neatly arranged

    **Paper Plates - Party Events Context:**
    1. **Party napkins** - Color-coordinated disposable napkins that complement paper plates for cohesive table settings
    2. **Plastic tablecloths** - Easy-cleanup table coverings that protect surfaces and match disposable plate themes
    3. **Serving utensils** - Disposable or reusable serving tools for transferring food onto paper plates
    4. **Beverage dispensers** - Large-capacity drink servers that complement disposable dinnerware for parties
    5. **Trash receptacles** - Additional waste containers needed for increased disposable plate and party cleanup
    6. **Cupcake stands** - Multi-tiered displays that enhance presentation of desserts served on paper plates
    7. **Party platters** - Disposable serving trays that hold appetizers and snacks alongside paper plates
    8. **Table centerpieces** - Decorative items that enhance the party atmosphere while using disposable dinnerware
    9. **Disposable cutlery sets** - Forks, knives, and spoons that match paper plates for complete dining solutions
    10. **Party favor bags** - Themed bags that complement paper plates and provide guests with take-home treats

    Requirements:
    - Each recommendation should be 2-4 words describing a specific product type
    - Focus on context-specific tools rather than generic items
    - Consider the scale/quantity of the {anchor_item} (individual vs. bulk size)
    - Think about the complete user experience in each context
    - Recommendations should be from non-grocery categories but directly support the anchor item's usage

    Avoid recommending:
    - Grocery items, food, beverages, or consumable products
    - Items that are not directly used with or related to the {anchor_item}
    - Generic household items that don't specifically enhance the anchor item experience
\end{lstlisting}
\end{tcolorbox}

\label{appendix:llm-prompts-theme-rec}
\begin{tcolorbox}[promptbox]
\begin{lstlisting}[backgroundcolor=\color{black!0}, basicstyle=\small\ttfamily\color{black}]
    For each recommendation, provide a brief explanation of:
    1. How it specifically enhances the {anchor_item} experience in that context
    2. What problem it solves or convenience it adds
    Output format: List of dictionaries, one for each context:
    [
    {{
        "context": "Context Name 1",
        "recs": ["Item Name 1", "Item Name 2", "Item Name 3", "Item Name 4", "Item Name 5", "Item Name 6", "Item Name 7", "Item Name 8", "Item Name 9", "Item Name 10"],
        "explanations": ["Explanation 1", "Explanation 2", "Explanation 3", "Explanation 4", "Explanation 5", "Explanation 6", "Explanation 7", "Explanation 8", "Explanation 9", "Explanation 10"]
    }},
    {{
        "context": "Context Name 2", 
        "recs": ["Item Name 1", "Item Name 2", "Item Name 3", "Item Name 4", "Item Name 5", "Item Name 6", "Item Name 7", "Item Name 8", "Item Name 9", "Item Name 10"],
        "explanations": ["Explanation 1", "Explanation 2", "Explanation 3", "Explanation 4", "Explanation 5", "Explanation 6", "Explanation 7", "Explanation 8", "Explanation 9", "Explanation 10"]
    }},
    {{
        "context": "Context Name 3",
        "recs": ["Item Name 1", "Item Name 2", "Item Name 3", "Item Name 4", "Item Name 5", "Item Name 6", "Item Name 7", "Item Name 8", "Item Name 9", "Item Name 10"], 
        "explanations": ["Explanation 1", "Explanation 2", "Explanation 3", "Explanation 4", "Explanation 5", "Explanation 6", "Explanation 7", "Explanation 8", "Explanation 9", "Explanation 10"]
    }},
    {{
        "context": "Context Name 4",
        "recs": ["Item Name 1", "Item Name 2", "Item Name 3", "Item Name 4", "Item Name 5", "Item Name 6", "Item Name 7", "Item Name 8", "Item Name 9", "Item Name 10"], 
        "explanations": ["Explanation 1", "Explanation 2", "Explanation 3", "Explanation 4", "Explanation 5", "Explanation 6", "Explanation 7", "Explanation 8", "Explanation 9", "Explanation 10"]
    }},
    {{
        "context": "Context Name 5",
        "recs": ["Item Name 1", "Item Name 2", "Item Name 3", "Item Name 4", "Item Name 5", "Item Name 6", "Item Name 7", "Item Name 8", "Item Name 9", "Item Name 10"], 
        "explanations": ["Explanation 1", "Explanation 2", "Explanation 3", "Explanation 4", "Explanation 5", "Explanation 6", "Explanation 7", "Explanation 8", "Explanation 9", "Explanation 10"]
    }}
    ]
\end{lstlisting}
\end{tcolorbox}

\subsection{LLM Candidate Evaluator}
\label{appendix:llm-prompts-gen_evaluator}
\begin{tcolorbox}[promptbox]
\begin{lstlisting}[backgroundcolor=\color{black!0}, basicstyle=\small\ttfamily\color{black}]
You are an e-commerce recommendation verification agent specializing in cross-category discovery.

TASK: Verify and score this single recommendation for the anchor item, focusing on cross-category discovery potential.

ANCHOR ITEM: {anchor_item}

RECOMMENDATION: {recommendation}
EXPLANATION: {explanation}

EVALUATION CRITERIA (prioritized for cross-category discovery):

1. CROSS-CATEGORY DISCOVERY POTENTIAL (40% weight):
   - Does this recommendation introduce customers to a different product category?
   - Will it expand their shopping scope beyond their original intent?
   - Does it create valuable cross-category connections?

2. CONTEXTUAL RELEVANCE (30% weight):
   - Does the recommendation make sense in the customer's broader context/lifestyle?
   - Are there logical use cases where both items would be valuable together?
   - Consider shared purposes, occasions, or user scenarios (not just direct functionality)

3. EXPLANATION QUALITY (20% weight):
   - Is the reasoning clear and logical?
   - Does it explain the cross-category connection effectively?
   - Does it help customers understand why this recommendation makes sense?

4. PURCHASE LIKELIHOOD (10% weight):
   - Would a customer interested in the anchor item reasonably consider this recommendation?
   - Does it align with customer behavior patterns?

SCORING GUIDELINES:
- HIGH SCORES (0.8-1.0): Strong cross-category discovery with clear contextual relevance
- MEDIUM SCORES (0.5-0.7): Some cross-category potential but weaker connections
- LOW SCORES (0.0-0.4): Poor cross-category discovery or weak relevance

IMPORTANT: 
- Prioritize broader lifestyle/contextual connections over narrow functional similarity
- Value recommendations that genuinely expand customer discovery
- Consider room enhancement, lifestyle compatibility, complementary use cases
- Don't penalize for being cross-category - that's the goal!

Provide a score from 0.0 to 1.0.
Return only valid JSON:

{{
  "score": <decimal 0.0 to 1.0>,
  "reasoning": "<brief explanation for the score, emphasizing cross-category discovery value>"
}}
\end{lstlisting}
\end{tcolorbox}

\subsection{Recommendation Item Evaluator}
\label{appendix:llm-prompts-2}
\begin{tcolorbox}[promptbox]
\begin{lstlisting}[backgroundcolor=\color{black!0}, basicstyle=\small\ttfamily\color{black}]
You are an e-commerce recommendation evaluation agent specializing in cross-category product discovery.
EVALUATION TASK:
Assess the quality of a cross-category recommendation and its matching accuracy.

INPUT DATA:
- Anchor Category: {anchor_pt}
- LLM Recommendation: {llm_rec}  
- Matched Product Category: {rec_pt}

EVALUATION CRITERIA:
Rate the recommendation on these four dimensions (0.0-1.0 scale):

1. CROSS-CATEGORY DISCOVERY (25%): Does the recommendation introduce customers to a meaningfully different but related category?
2. RELEVANCE & COHERENCE (35%): Is there a logical connection between the anchor category and recommendation that customers would understand?
3. PRACTICAL UTILITY (25%): Would customers realistically consider purchasing both categories together or in sequence?
4. MATCHING ACCURACY (15%): Does the matched product category align well with the LLM's intended recommendation?

SCORING GUIDELINES:
- 0.8-1.0: Excellent - Strong cross-category connection with high purchase likelihood
- 0.6-0.79: Good - Clear relevance with moderate cross-category value  
- 0.4-0.59: Average - Some connection exists but limited discovery potential
- 0.2-0.39: Poor - Weak connection or minimal cross-category benefit
- 0.0-0.19: Very Poor - No meaningful relationship or completely unrelated

DOMAIN DEFINITIONS:
Consider these as unified domains when evaluating connections:
- Beauty & Personal Care: makeup, skincare, hair care, fragrances, personal hygiene
- Food & Kitchen: cooking, baking, food storage, kitchen tools, dining
- Health & Wellness: supplements, medications, fitness, medical devices
- Home & Living: decor, furniture, cleaning, organization, textiles
- Family & Parenting: baby care, toys, family activities, child safety
- Pets: pet food, toys, grooming, health, accessories

IMPORTANT NOTES:
- Assume the LLM recommendation itself is valid and appropriate
- Focus on whether it achieves cross-category discovery goals
- Be generous with connections within the same broad domain
- Only score below 0.4 if categories serve completely different purposes or customer needs

OUTPUT FORMAT:
Return only valid JSON:
{{
    "score": <decimal from 0.0 to 1.0>,
    "reasoning": "<concise explanation covering relevance, cross-category value, and matching accuracy>"
}}
\end{lstlisting}
\end{tcolorbox}

\section{Sample LLM Outputs}
\label{appendix:llm-outputs}
Here are some sample LLM outputs (with the format slightly modified for easier reading) for couple of top grocery anchors
\begin{tcolorbox}[promptbox]
\begin{lstlisting}[backgroundcolor=\color{black!0}, basicstyle=\small\ttfamily\color{black}]
Input: "Eggs (fresh), ProductType: Eggs"

Output: "recs": [
    "Egg Poachers: These products offer various ways to cook eggs with convenience and ease, catering to the versatile nature of eggs",
    "Egg Cookers: Egg cookers provide a fuss-free way of cooking eggs with consistency and without the need for a stove",
    "Egg Slicers: Egg slicers provide a quick and easy way to slice eggs for garnishes or salads",
    "Egg Cups: Egg cups provide a classy way of serving a boiled egg, commonly used for breakfast or brunch",
    "Egg Separators: Egg separators provide an effortless way of separating egg yolk from egg white",
    "Egg Timers: Egg timers ensure that you cook your eggs to your desired level of doneness, preventing overcooking",
    "Egg Beaters: Egg beaters help in creating fluffy and light egg mixtures for baking and cooking",
    "Egg Rings: These offer a simple way of creating uniform egg patties for breakfast sandwiches",
    "Egg Boilers: Egg boilers eliminate the guesswork in boiling eggs, producing consistent results every time",
    "Egg Topper: It's is a nifty gadget that takes the top off boiled eggs, allowing for easy eating",
    "Eggshell Remover: Eggshell remover is a handy tool that helps in removing eggshells that are stubbornly stuck to the egg white",
    "Egg Whisk: Egg whisk is a versatile tool that can be used for whisking egg whites, making mayonnaise, and emulsifying",
    "Egg Separator Strainer: Egg separator strainer is an all-in-one tool that separates the egg yolk from the egg white and strains the egg white for a smoother consistency"
  ]
\end{lstlisting}
\end{tcolorbox}

\begin{tcolorbox}[promptbox]
\begin{lstlisting}[backgroundcolor=\color{black!0}, basicstyle=\small\ttfamily\color{black}]
Input: "Dry Dog Foods, ProductType: Dog Food"

Output: "recs": [
    "Dog bowls: Dog bowls will enhance the feeding experience for dogs and make sure food doesn't get spilled.",
    "Dog Food Mats: Dog food mats catch any spills from feeding and keep the surrounding area clean",
    "Automatic Pet Feeders: To feed your pet on a schedule without any hassle.",
    "Chew Toys: Chew toys help to keep your puppy engaged and entertained.",
    "Dog Treat Jars: Dog treat jars can be used to store and serve delicious treats alongside the usual dry dog food.",
    "Dog Water Dispensers: Dog water dispensers can be used to supply clean and fresh drinking water along with the dry dog food.",
    "Dog Food Scoops: Dog food scoops help properly portion meals and serve the correct amount of food to the dog.",
    "Dog Food Containers: Dog food containers can be used to store and pack dry food to ensure its freshness and retain nutritional value.",
    "Dog Food Trays: Dog food trays help to portion and serve the dry dog food in a visually appealing way",
    "Dog Food Bag Clips: Dog food bag clips ensure dry dog food stays fresh for a longer duration"
    ...
  ]
\end{lstlisting}
\end{tcolorbox}

\begin{table}[h]
    \centering
    \caption{Top 3 and Bottom 3 Item Types by CE Score Performance Across All Categories}
    \label{tab:ce_mean_comprehensive}
    \resizebox{\columnwidth}{!}{%
    \begin{tabular}{c l >{\columncolor{yellow!20}}c c c c c c}
    \toprule
    \textbf{Product Category} & \textbf{Item Type} & \textbf{CE Mean} & \textbf{LLM Mean} & \textbf{Combined Mean} & \textbf{Pct Poor CE} & \textbf{Pct Poor LLM} & \textbf{Pct Poor Combined} \\
    \midrule
    \multirow{3}{*}{Baby} & \cellcolor{green!20} Disposable Diaper Bags & \textbf{0.9686} & 0.7014 & 0.6789 & 6.97 & 7.38 & 11.89 \\
    & Baby Wipes & 0.9678 & 0.6924 & 0.6698 & 4.95 & 15.39 & 12.78 \\
    & Toilet Training Pants & 0.9658 & 0.7008 & 0.6775 & 5.78 & 14.50 & 12.44 \\
    \cmidrule(l){2-8}
    & \cellcolor{red!20}\textit{Diaper Pail Liners} & \textit{0.8694} & 0.6812 & 0.5923 & 24.10 & 21.01 & 24.92 \\
    & Nursing Pads & 0.9221 & 0.6852 & 0.6347 & 11.89 & 17.08 & 16.94 \\
    & Baby Formula & 0.9293 & 0.6806 & 0.6332 & 12.09 & 17.28 & 17.95 \\
    \midrule
    \multirow{3}{*}{Food \& Beverages} & \cellcolor{green!20}\textbf{Worcestershire Sauce} & \textbf{0.9986} & 0.6693 & 0.6682 & 0.41 & 18.85 & 18.44 \\
    & Mustards & 0.9959 & 0.7008 & 0.6982 & 1.22 & 13.41 & 6.71 \\
    & Ground Coffee & 0.9946 & 0.6931 & 0.6894 & 1.20 & 16.67 & 14.66 \\
    \cmidrule(l){2-8}
    & \cellcolor{red!20}\textit{Clams} & \textit{0.8852} & 0.6795 & 0.5938 & 13.52 & 16.80 & 25.00 \\
    & English Muffins & 0.9085 & 0.7043 & 0.6470 & 12.30 & 8.61 & 12.70 \\
    & Waffles & 0.9165 & 0.6908 & 0.6372 & 11.17 & 18.97 & 19.29 \\
    \midrule
    \multirow{3}{*}{Home \& Garden} & \cellcolor{green!20}\textbf{Straws} & \textbf{0.9906} & 0.6689 & 0.6617 & 2.81 & 23.90 & 17.87 \\
    & Parchment Paper & 0.9883 & 0.6987 & 0.6905 & 2.58 & 12.66 & 6.99 \\
    & Napkins & 0.9868 & 0.6785 & 0.6697 & 2.77 & 21.12 & 15.90 \\
    \cmidrule(l){2-8}
    & \cellcolor{red!20}\textit{Furnace Filters} & \textit{0.8743} & 0.6687 & 0.5869 & 16.08 & 20.32 & 24.17 \\
    & Drain Openers & 0.8973 & 0.6693 & 0.6016 & 14.85 & 21.52 & 19.66 \\
    & Pool Chemicals & 0.9108 & 0.6579 & 0.5987 & 12.90 & 20.93 & 22.24 \\
    \midrule
    \multirow{3}{*}{Pet Supplies} & \cellcolor{green!20}\textbf{Pet Grooming Wipes} & \textbf{0.9705} & 0.6962 & 0.6746 & 7.23 & 13.65 & 10.44 \\
    & Animal Dewormers & 0.9505 & 0.6860 & 0.6529 & 6.89 & 19.19 & 12.03 \\
    & Pet Milk Replacers & 0.9446 & 0.6722 & 0.6378 & 7.91 & 18.05 & 15.82 \\
    \cmidrule(l){2-8}
    & \cellcolor{red!20}\textit{Bird Food} & \textit{0.8188} & 0.6527 & 0.5198 & 23.65 & 23.24 & 31.95 \\
    & Fish Food & 0.8327 & 0.6824 & 0.5635 & 19.06 & 20.29 & 28.48 \\
    & Small Animal Food & 0.8523 & 0.6688 & 0.5792 & 18.52 & 20.01 & 24.49 \\
    \bottomrule
    \end{tabular}%
    }
\end{table}

\begin{table}[h]
    \centering
    \caption{Top 3 and Bottom 3 Item Types by LLM Score Across All Categories}
    \label{tab:llm_mean_comprehensive}
    \resizebox{\columnwidth}{!}{%
    \begin{tabular}{c l c >{\columncolor{yellow!20}}c c c c c}
    \toprule
    \textbf{Product Category} & \textbf{Item Type} & \textbf{CE Mean} & \textbf{LLM Mean} & \textbf{Combined Mean} & \textbf{Pct Poor CE} & \textbf{Pct Poor LLM} & \textbf{Pct Poor Combined} \\
    \midrule
    \multirow{3}{*}{Baby} & \cellcolor{green!20}\textbf{Baby Bottle Nipples} & 0.9622 & \textbf{0.7090} & 0.6841 & 7.14 & 13.17 & 11.69 \\
    & Disposable Baby Diapers & 0.9543 & 0.7073 & 0.6754 & 8.14 & 11.98 & 11.61 \\
    & Baby Cereals & 0.9447 & 0.7014 & 0.6626 & 10.25 & 13.87 & 12.37 \\
    \cmidrule(l){2-8}
    & \cellcolor{red!20}\textit{Pediatric Nutrition Shakes} & 0.9323 & \textit{0.6721} & 0.6286 & 9.97 & 19.89 & 17.85 \\
    & Baby Foods & 0.9432 & 0.6762 & 0.6380 & 8.93 & 18.17 & 17.56 \\
    & Baby Formula & 0.9293 & 0.6806 & 0.6332 & 12.09 & 17.28 & 17.95 \\
    \midrule
    \multirow{3}{*}{Food \& Beverages} & \cellcolor{green!20}\textbf{Shrimp \& Prawns} & 0.9774 & \textbf{0.7319} & 0.7167 & 3.55 & 9.12 & 7.29 \\
    & Mushrooms & 0.9844 & 0.7235 & 0.7125 & 3.26 & 9.05 & 6.71 \\
    & Olives & 0.9834 & 0.7225 & 0.7100 & 2.90 & 4.56 & 4.36 \\
    \cmidrule(l){2-8}
    & \cellcolor{red!20}\textit{Applesauce \& Fruit Sauces} & 0.9518 & \textit{0.6225} & 0.5962 & 7.85 & 33.06 & 25.21 \\
    & Pretzels & 0.9851 & 0.6450 & 0.6348 & 3.66 & 24.39 & 19.31 \\
    & Toaster Pastries & 0.9763 & 0.6528 & 0.6359 & 4.07 & 25.30 & 21.04 \\
    \midrule
    \multirow{3}{*}{Home \& Garden} & \cellcolor{green!20}\textbf{Fabric Starch} & 0.9676 & \textbf{0.7267} & 0.7026 & 4.86 & 3.24 & 6.48 \\
    & Carpet Cleaners & 0.9724 & 0.7205 & 0.7010 & 5.83 & 7.19 & 5.97 \\
    & Fabric Softeners & 0.9653 & 0.7166 & 0.6917 & 6.52 & 11.62 & 9.26 \\
    \cmidrule(l){2-8}
    & \cellcolor{red!20}\textit{Matches} & 0.9368 & \textit{0.6317} & 0.6052 & 12.90 & 31.05 & 29.03 \\
    & Pet Stain Remover & 0.9812 & 0.6444 & 0.6328 & 4.44 & 25.40 & 15.32 \\
    & Household Cleaners & 0.9651 & 0.6498 & 0.6273 & 6.00 & 25.12 & 20.94 \\
    \midrule
    \multirow{3}{*}{Pet Supplies} & \cellcolor{green!20}\textbf{Dog Waste Bags} & 0.9441 & \textbf{0.7057} & 0.6706 & 8.45 & 16.08 & 12.81 \\
    & Pet Grooming Wipes & 0.9705 & 0.6962 & 0.6746 & 7.23 & 13.65 & 10.44 \\
    & Pet Training Pads & 0.9396 & 0.6900 & 0.6514 & 8.48 & 17.88 & 17.17 \\
    \cmidrule(l){2-8}
    & \cellcolor{red!20}\textit{Pet Relaxants} & 0.9153 & \textit{0.5919} & 0.5414 & 9.27 & 34.27 & 33.06 \\
    & Pet Diapers & 0.9372 & 0.6408 & 0.5956 & 8.20 & 26.64 & 29.10 \\
    & Cat Litter & 0.9231 & 0.6507 & 0.6029 & 14.10 & 25.46 & 24.42 \\
    \bottomrule
    \end{tabular}%
    }
\end{table}

\begin{table}[h]
    \centering
    \caption{Top 3 and Bottom 3 Item Types by Combined Score Across All Categories}
    \label{tab:combined_mean_comprehensive}
    \resizebox{\columnwidth}{!}{%
    \begin{tabular}{c l c c >{\columncolor{yellow!20}}c c c c}
    \toprule
    \textbf{Product Category} & \textbf{Item Type} & \textbf{CE Mean} & \textbf{LLM Mean} & \textbf{Combined Mean} & \textbf{Pct Poor CE} & \textbf{Pct Poor LLM} & \textbf{Pct Poor Combined} \\
    \midrule
    \multirow{3}{*}{Baby} & \cellcolor{green!20}\textbf{Baby Bottle Nipples} & 0.9622 & 0.7090 & \textbf{0.6841} & 7.14 & 13.17 & 11.69 \\
    & Disposable Diaper Bags & 0.9686 & 0.7014 & 0.6789 & 6.97 & 7.38 & 11.89 \\
    & Toilet Training Pants & 0.9658 & 0.7008 & 0.6775 & 5.78 & 14.50 & 12.44 \\
    \cmidrule(l){2-8}
    & \cellcolor{red!20}\textit{Diaper Pail Liners} & 0.8694 & 0.6812 & \textit{0.5923} & 24.10 & 21.01 & 24.92 \\
    & Pediatric Nutrition Shakes & 0.9323 & 0.6721 & 0.6286 & 9.97 & 19.89 & 17.85 \\
    & Baby Formula & 0.9293 & 0.6806 & 0.6332 & 12.09 & 17.28 & 17.95 \\
    \midrule
    \multirow{3}{*}{Food \& Beverages} & \cellcolor{green!20}\textbf{Shrimp \& Prawns} & 0.9774 & 0.7319 & \textbf{0.7167} & 3.55 & 9.12 & 7.29 \\
    & Mushrooms & 0.9844 & 0.7235 & 0.7125 & 3.26 & 9.05 & 6.71 \\
    & Evaporated Milks & 0.9853 & 0.7193 & 0.7106 & 3.21 & 6.02 & 6.02 \\
    \cmidrule(l){2-8}
    & \cellcolor{red!20}\textit{Clams} & 0.8852 & 0.6795 & \textit{0.5938} & 13.52 & 16.80 & 25.00 \\
    & Applesauce \& Fruit Sauces & 0.9518 & 0.6225 & 0.5962 & 7.85 & 33.06 & 25.21 \\
    & Pancakes & 0.9292 & 0.6716 & 0.6233 & 10.93 & 19.67 & 17.38 \\
    \midrule
    \multirow{3}{*}{Home \& Garden} & \cellcolor{green!20}\textbf{Fabric Starch} & 0.9676 & 0.7267 & \textbf{0.7026} & 4.86 & 3.24 & 6.48 \\
    & Carpet Cleaners & 0.9724 & 0.7205 & 0.7010 & 5.83 & 7.19 & 5.97 \\
    & Paper Plates & 0.9822 & 0.7092 & 0.6969 & 3.73 & 12.93 & 9.76 \\
    \cmidrule(l){2-8}
    & \cellcolor{red!20}\textit{Furnace Filters} & 0.8743 & 0.6687 & \textit{0.5869} & 16.08 & 20.32 & 24.17 \\
    & Pool Chemicals & 0.9108 & 0.6579 & 0.5987 & 12.90 & 20.93 & 22.24 \\
    & Drain Openers & 0.8973 & 0.6693 & 0.6016 & 14.85 & 21.52 & 19.66 \\
    \midrule
    \multirow{3}{*}{Pet Supplies} & \cellcolor{green!20}\textbf{Pet Grooming Wipes} & 0.9705 & 0.6962 & \textbf{0.6746} & 7.23 & 13.65 & 10.44 \\
    & Dog Waste Bags & 0.9441 & 0.7057 & 0.6706 & 8.45 & 16.08 & 12.81 \\
    & Animal Dewormers & 0.9505 & 0.6860 & 0.6529 & 6.89 & 19.19 & 12.03 \\
    \cmidrule(l){2-8}
    & \cellcolor{red!20}\textit{Bird Food} & 0.8188 & 0.6527 & \textit{0.5198} & 23.65 & 23.24 & 31.95 \\
    & Pet Relaxants & 0.9153 & 0.5919 & 0.5414 & 9.27 & 34.27 & 33.06 \\
    & Fish Food & 0.8327 & 0.6824 & 0.5635 & 19.06 & 20.29 & 28.48 \\
    \bottomrule
    \end{tabular}%
    }
\end{table}

\begin{table}[h]
    \centering
    \caption{Top 3 and Bottom 3 Item Types by PCT\_POOR\_CE Performance Across All Categories}
    \label{tab:pct_poor_ce_comprehensive}
    \resizebox{\columnwidth}{!}{%
    \begin{tabular}{c l c c c >{\columncolor{yellow!20}}c c c}
    \toprule
    \textbf{Product Category} & \textbf{Item Type} & \textbf{CE Mean} & \textbf{LLM Mean} & \textbf{Combined Mean} & \textbf{Pct Poor CE} & \textbf{Pct Poor LLM} & \textbf{Pct Poor Combined} \\
    \midrule
    \multirow{3}{*}{Baby} & \cellcolor{green!20}\textbf{Diaper Pail Liners} & 0.8694 & 0.6812 & 0.5923 & \textbf{24.10} & 21.01 & 24.92 \\
    & Baby Formula & 0.9293 & 0.6806 & 0.6332 & 12.09 & 17.28 & 17.95 \\
    & Nursing Pads & 0.9221 & 0.6852 & 0.6347 & 11.89 & 17.08 & 16.94 \\
    \cmidrule(l){2-8}
    & \cellcolor{red!20}\textit{Baby Wipes} & 0.9678 & 0.6924 & 0.6698 & \textit{4.95} & 15.39 & 12.78 \\
    & Toilet Training Pants & 0.9658 & 0.7008 & 0.6775 & 5.78 & 14.50 & 12.44 \\
    & Baby Snacks & 0.9624 & 0.6932 & 0.6678 & 6.38 & 14.35 & 11.51 \\
    \midrule
    \multirow{3}{*}{Food \& Beverages} & \cellcolor{green!20}\textbf{Clams} & 0.8852 & 0.6795 & 0.5938 & \textbf{13.52} & 16.80 & 25.00 \\
    & English Muffins & 0.9085 & 0.7043 & 0.6470 & 12.30 & 8.61 & 12.70 \\
    & Waffles & 0.9165 & 0.6908 & 0.6372 & 11.17 & 18.97 & 19.29 \\
    \cmidrule(l){2-8}
    & \cellcolor{red!20}\textit{Worcestershire Sauce} & 0.9986 & 0.6693 & 0.6682 & \textit{0.41} & 18.85 & 18.44 \\
    & Ground Coffee & 0.9946 & 0.6931 & 0.6894 & 1.20 & 16.67 & 14.66 \\
    & Meal Kits & 0.9933 & 0.6647 & 0.6605 & 1.21 & 25.10 & 18.22 \\
    \midrule
    \multirow{3}{*}{Home \& Garden} & \cellcolor{green!20}\textbf{Furnace Filters} & 0.8743 & 0.6687 & 0.5869 & \textbf{16.08} & 20.32 & 24.17 \\
    & Drain Openers & 0.8973 & 0.6693 & 0.6016 & 14.85 & 21.52 & 19.66 \\
    & Laundry Sanitizers & 0.9380 & 0.6875 & 0.6468 & 13.64 & 16.56 & 14.85 \\
    \cmidrule(l){2-8}
    & \cellcolor{red!20}\textit{Parchment Paper} & 0.9883 & 0.6987 & 0.6905 & \textit{2.58} & 12.66 & 6.99 \\
    & Napkins & 0.9868 & 0.6785 & 0.6697 & 2.77 & 21.12 & 15.90 \\
    & Straws & 0.9906 & 0.6689 & 0.6617 & 2.81 & 23.90 & 17.87 \\
    \midrule
    \multirow{3}{*}{Pet Supplies} & \cellcolor{green!20}\textbf{Bird Food} & 0.8188 & 0.6527 & 0.5198 & \textbf{23.65} & 23.24 & 31.95 \\
    & Fish Food & 0.8327 & 0.6824 & 0.5635 & 19.06 & 20.29 & 28.48 \\
    & Small Animal Food & 0.8523 & 0.6688 & 0.5792 & 18.52 & 20.01 & 24.49 \\
    \cmidrule(l){2-8}
    & \cellcolor{red!20}\textit{Animal Dewormers} & 0.9505 & 0.6860 & 0.6529 & \textit{6.89} & 19.19 & 12.03 \\
    & Pet Grooming Wipes & 0.9705 & 0.6962 & 0.6746 & 7.23 & 13.65 & 10.44 \\
    & Pet Milk Replacers & 0.9446 & 0.6722 & 0.6378 & 7.91 & 18.05 & 15.82 \\
    \bottomrule
    \end{tabular}%
    }
\end{table}

\begin{table}[h]
    \centering
    \caption{Top 3 and Bottom 3 Item Types by PCT\_POOR\_LLM Performance Across All Categories}
    \label{tab:pct_poor_llm_comprehensive}
    \resizebox{\columnwidth}{!}{%
    \begin{tabular}{c l c c c c >{\columncolor{yellow!20}}c c}
    \toprule
    \textbf{Product Category} & \textbf{Item Type} & \textbf{CE Mean} & \textbf{LLM Mean} & \textbf{Combined Mean} & \textbf{Pct Poor CE} & \textbf{Pct Poor LLM} & \textbf{Pct Poor Combined} \\
    \midrule
    \multirow{3}{*}{Baby} & \cellcolor{green!20}\textbf{Diaper Pail Liners} & 0.8694 & 0.6812 & 0.5923 & 24.10 & \textbf{21.01} & 24.92 \\
    & Pediatric Nutrition Shakes & 0.9323 & 0.6721 & 0.6286 & 9.97 & 19.89 & 17.85 \\
    & Baby Juices & 0.9569 & 0.6854 & 0.6574 & 7.03 & 18.33 & 14.87 \\
    \cmidrule(l){2-8}
    & \cellcolor{red!20}\textit{Disposable Diaper Bags} & 0.9686 & 0.7014 & 0.6789 & 6.97 & \textit{7.38} & 11.89 \\
    & Disposable Baby Diapers & 0.9543 & 0.7073 & 0.6754 & 8.14 & 11.98 & 11.61 \\
    & Baby Bottle Nipples & 0.9622 & 0.7090 & 0.6841 & 7.14 & 13.17 & 11.69 \\
    \midrule
    \multirow{3}{*}{Food \& Beverages} & \cellcolor{green!20}\textbf{Applesauce \& Fruit Sauces} & 0.9518 & 0.6225 & 0.5962 & 7.85 & \textbf{33.06} & 25.21 \\
    & Sandwiches, Filled Rolls \& Wraps & 0.9712 & 0.6551 & 0.6356 & 4.55 & 25.69 & 19.56 \\
    & Candy Bars \& Miniatures & 0.9904 & 0.6626 & 0.6570 & 1.23 & 25.51 & 13.17 \\
    \cmidrule(l){2-8}
    & \cellcolor{red!20}\textit{Olives} & 0.9834 & 0.7225 & 0.7100 & 2.90 & \textit{4.56} & 4.36 \\
    & Meatballs & 0.9781 & 0.7094 & 0.6952 & 3.69 & 5.74 & 7.79 \\
    & Evaporated Milks & 0.9853 & 0.7193 & 0.7106 & 3.21 & 6.02 & 6.02 \\
    \midrule
    \multirow{3}{*}{Home \& Garden} & \cellcolor{green!20}\textbf{Matches} & 0.9368 & 0.6317 & 0.6052 & 12.90 & \textbf{31.05} & 29.03 \\
    & Dusters & 0.9635 & 0.6524 & 0.6271 & 5.74 & 25.82 & 16.70 \\
    & Pet Stain Remover & 0.9812 & 0.6444 & 0.6328 & 4.44 & 25.40 & 15.32 \\
    \cmidrule(l){2-8}
    & \cellcolor{red!20}\textit{Fabric Starch} & 0.9676 & 0.7267 & 0.7026 & 4.86 & \textit{3.24} & 6.48 \\
    & Carpet Cleaners & 0.9724 & 0.7205 & 0.7010 & 5.83 & 7.19 & 5.97 \\
    & Food Wraps & 0.9782 & 0.7050 & 0.6906 & 3.57 & 10.59 & 8.11 \\
    \midrule
    \multirow{3}{*}{Pet Supplies} & \cellcolor{green!20}\textbf{Pet Relaxants} & 0.9153 & 0.5919 & 0.5414 & 9.27 & \textbf{34.27} & 33.06 \\
    & Pet Diapers & 0.9372 & 0.6408 & 0.5956 & 8.20 & 26.64 & 29.10 \\
    & Cat Litter & 0.9231 & 0.6507 & 0.6029 & 14.10 & 25.46 & 24.42 \\
    \cmidrule(l){2-8}
    & \cellcolor{red!20}\textit{Pet Grooming Wipes} & 0.9705 & 0.6962 & 0.6746 & 7.23 & \textit{13.65} & 10.44 \\
    & Pet Itch Remedies & 0.9113 & 0.6753 & 0.6176 & 12.90 & 14.52 & 14.92 \\
    & Cat Litter Box Liners & 0.9314 & 0.6877 & 0.6408 & 14.27 & 15.07 & 15.34 \\
    \bottomrule
    \end{tabular}%
    }
\end{table}

\begin{table}[h]
    \centering
    \caption{Top 3 and Bottom 3 Item Types by PCT\_POOR\_COMBINED Performance Across All Categories}
    \label{tab:pct_poor_combined_comprehensive}
    \resizebox{\columnwidth}{!}{%
    \begin{tabular}{c l c c c c c >{\columncolor{yellow!20}}c}
    \toprule
    \textbf{Product Category} & \textbf{Item Type} & \textbf{CE Mean} & \textbf{LLM Mean} & \textbf{Combined Mean} & \textbf{Pct Poor CE} & \textbf{Pct Poor LLM} & \textbf{Pct Poor Combined} \\
    \midrule
    \multirow{3}{*}{Baby} & \cellcolor{green!20}\textbf{Diaper Pail Liners} & 0.8694 & 0.6812 & 0.5923 & 24.10 & 21.01 & \textbf{24.92} \\
    & Nipple Soothing Creams & 0.9465 & 0.6967 & 0.6570 & 7.52 & 16.67 & 18.29 \\
    & Baby Formula & 0.9293 & 0.6806 & 0.6332 & 12.09 & 17.28 & 17.95 \\
    \cmidrule(l){2-8}
    & \cellcolor{red!20}\textit{Baby Snacks} & 0.9624 & 0.6932 & 0.6678 & 6.38 & 14.35 & \textit{11.51} \\
    & Disposable Baby Diapers & 0.9543 & 0.7073 & 0.6754 & 8.14 & 11.98 & 11.61 \\
    & Baby Bottle Nipples & 0.9622 & 0.7090 & 0.6841 & 7.14 & 13.17 & 11.69 \\
    \midrule
    \multirow{3}{*}{Food \& Beverages} & \cellcolor{green!20}\textbf{Applesauce \& Fruit Sauces} & 0.9518 & 0.6225 & 0.5962 & 7.85 & 33.06 & \textbf{25.21} \\
    & Clams & 0.8852 & 0.6795 & 0.5938 & 13.52 & 16.80 & 25.00 \\
    & Hot Sauces & 0.9460 & 0.6674 & 0.6314 & 8.91 & 17.81 & 22.67 \\
    \cmidrule(l){2-8}
    & \cellcolor{red!20}\textit{Jerky \& Dried Meats} & 0.9807 & 0.7000 & 0.6889 & 2.89 & 13.22 & \textit{4.13} \\
    & Olives & 0.9834 & 0.7225 & 0.7100 & 2.90 & 4.56 & 4.36 \\
    & Watermelons & 0.9888 & 0.7150 & 0.7078 & 2.45 & 8.49 & 4.40 \\
    \midrule
    \multirow{3}{*}{Home \& Garden} & \cellcolor{green!20}\textbf{Matches} & 0.9368 & 0.6317 & 0.6052 & 12.90 & 31.05 & \textbf{29.03} \\
    & Furnace Filters & 0.8743 & 0.6687 & 0.5869 & 16.08 & 20.32 & 24.17 \\
    & Pool Chemicals & 0.9108 & 0.6579 & 0.5987 & 12.90 & 20.93 & 22.24 \\
    \cmidrule(l){2-8}
    & \cellcolor{red!20}\textit{Carpet Cleaners} & 0.9724 & 0.7205 & 0.7010 & 5.83 & 7.19 & \textit{5.97} \\
    & Mulches & 0.9520 & 0.7080 & 0.6841 & 7.71 & 12.98 & 6.09 \\
    & Fabric Starch & 0.9676 & 0.7267 & 0.7026 & 4.86 & 3.24 & 6.48 \\
    \midrule
    \multirow{3}{*}{Pet Supplies} & \cellcolor{green!20}\textbf{Pet Relaxants} & 0.9153 & 0.5919 & 0.5414 & 9.27 & 34.27 & \textbf{33.06} \\
    & Bird Food & 0.8188 & 0.6527 & 0.5198 & 23.65 & 23.24 & 31.95 \\
    & Pet Diapers & 0.9372 & 0.6408 & 0.5956 & 8.20 & 26.64 & 29.10 \\
    \cmidrule(l){2-8}
    & \cellcolor{red!20}\textit{Pet Grooming Wipes} & 0.9705 & 0.6962 & 0.6746 & 7.23 & 13.65 & \textit{10.44} \\
    & Animal Dewormers & 0.9505 & 0.6860 & 0.6529 & 6.89 & 19.19 & 12.03 \\
    & Dog Waste Bags & 0.9441 & 0.7057 & 0.6706 & 8.45 & 16.08 & 12.81 \\
    \bottomrule
    \end{tabular}%
    }
\end{table}

\clearpage

\newpage

\section{Cart Ranker Feature Embeddings and Model Complexity}
\label{appendix:cart-transformer-complexity}
We represent categorical features such as platform and OG/GM type using embedding lookup tables. For price features, we apply a logarithmic transformation to handle the skewed distribution, then project these transformed values into a dense representation.
Customer features comprise of a total of 64 behavioral persona scores between 0 to 1, like cooking propensity, gender, new parent, pet ownership etc. These features are projected into a 128-dimensional dense representation.
The model architecture incorporates a four-layer transformer with 4 multi headed attention heads per layer for cart encoding and a single-layer cross-attention mechanism, totally approximately 1.2M trainable parameters. This efficient design enables real-time inference in our large-scale production environment while maintaining model expressiveness.

\section{Ablation Study with cart size}
Our best performing model uses up-to 50 items for cart context. Upon limiting the cart context to 30 items, we see a 0.7\% drop in NDCG@2 and 0.39\% drop in NDCG@4 metrics. We also analyze the lift in the metrics for different cart sizes in Table \ref{tab: ndcg gain cart size}. We see a higher lift in the NDCG metrics for larger carts compared to carts with fewer than 10 items. This showcases the model's capability in handling longer cart contexts.

\begin{table}[h]
  \caption{Comparison of NDCG Lift across different cart sizes}
  \label{tab: ndcg gain cart size}
  \centering
  \begin{tabular}{l c c c}
    \toprule
    \textbf{Cart Size} & \multicolumn{3}{c}{\textbf{NDCG Lift \%}} \\
    \cmidrule(l){2-4}
    & \textbf{@2} & \textbf{@4} & \textbf{@6} \\
    \midrule
    <10 & +22.96 & +23.70 & +13.03\\
    10-20& +28.21 & +27.72 & +15.93 \\
    20-30 & +29.49 & +28.42 & +16.72 \\
    30-40 & +30.56 & +28.13 & +17.02 \\
    40-50 & +30.20 & +27.75 & +17.02\\
    \bottomrule
  \end{tabular}
\end{table}

\end{document}